\documentclass[a4paper,fleqn,usenatbib]{mn2e}

\usepackage{aas_macros}

\usepackage[T1]{fontenc}
\usepackage{ae,aecompl}

\usepackage{graphicx}	
\usepackage{amsmath}	
\usepackage{amssymb}	

\newcommand{\hii}{H~{\sc ii}}

\title[Initial models and variable accretion rates]
{Impact of initial models and variable accretion rates on the pre-main-sequence evolution of massive and intermediate-mass stars
and the early evolution of \hii\ regions}

\author[L. Haemmerl\'e \& T. Peters]{
Lionel Haemmerl\'e$^{1}$\thanks{E-mail: lionel.haemmerle@unige.ch (LH)}
\& Thomas Peters$^{2}$
\\
$^{1}$Universit\"{a}t Heidelberg, Zentrum f\"{u}r Astronomie, Institut f\"{u}r Theoretische Astrophysik, Albert-Ueberle-Str. 2, D-69120 Heidelberg, Germany\\
$^{2}$Max-Planck-Institut f\"{u}r Astrophysik, Karl-Schwarzschild-Str. 1, D-85748 Garching, Germany
}

\date{Accepted XXX. Received YYY; in original form ZZZ}

\pubyear{2015}

\begin{document}
\label{firstpage}
\pagerange{\pageref{firstpage}--\pageref{lastpage}}
\maketitle

\begin{abstract}
Massive star formation requires the accretion of gas at high rate while the star is already bright.
Its actual luminosity depends sensitively on the stellar structure.
We compute pre-main-sequence tracks for massive and intermediate-mass stars with variable accretion rates
and study the evolution of stellar radius, effective temperature and ionizing luminosity,
starting at $2\,M_\odot$ with convective or radiative structures.
The radiative case shows a much stronger swelling of the protostar for high accretion rates than the convective case.
For radiative structures, the star is very sensitive to the accretion rate and reacts quickly to accretion bursts,
leading to considerable changes in photospheric properties on timescales as short as 100 -- 1000 yr.
The evolution for convective structures is much less influenced by the instantaneous accretion rate,
and produces a monotonically increasing ionizing flux that can be many orders of magnitude smaller than in the radiative case.
For massive stars, it results in a delay of the \hii\ region expansion by up to 10,000 yr.
In the radiative case, the \hii\ region can potentially be engulfed by the star during the swelling,
which never happens in the convective case.
We conclude that the early stellar structure has a large impact on the radiative feedback during the pre-main-sequence evolution
of massive protostars and introduces an important uncertainty that should be taken into account.
Because of their lower effective temperatures, our convective models may hint at a solution to an observed discrepancy
between the luminosity distribution functions of massive young stellar objects and compact \hii\ regions.
\end{abstract}




\section{Introduction}
\label{sec-intro}

Massive stars govern their environment through their various feedbacks: dynamical, chemical and radiative.
In particular, their strong UV radiation field is able to ionize large regions of gas in their surrounding.
It is expected that such an \hii\ region forms around a star as soon as its mass exceeds $\sim10\,M_\odot$ \citep{stahler2004}.
Due to the high pressure of the ionized gas, it was expected in the past that the growth of such an \hii\ region could reverse the accretion flow
and prevent further accretion, fixing the upper stellar mass limit at $\sim25\,M_\odot$ \citep{larson1971}.
However, recent three-dimensional simulations including ionizing feedback \citep{peters2010a,peters2010b,peters2010c,peters2011}
have shown that accretion is not halted by the high pressure of this hot ionized gas.
Indeed, it appears in these simulations that \hii\ regions show significant time variability and geometrical irregularity,
that allow the accretion flow to survive the increase in the gas pressure in the ionized regions.
Interestingly, this time variability appears to be in good agreement with observations
\citep{hughes1988,francohernandez2004,rodriguez2007,galvanmadrid2008,gomez2008,depree2014,depree2015,rivilla2015}.
Such variations may result from various causes, some related to the accretion flow itself
\citep{peters2010a,peters2010b,galvanmadrid2011,klassen2012a},
some related to the variations in the radiation field of the forming star \citep{klassen2012b}.

The pre-main-sequence (MS) evolution of massive stars remains an open issue.
The models of \cite{bernasconi1996a} and \cite{norberg2000},
computed with accretion rates in the range $10^{-5}-10^{-4}\,M_\odot\rm\,yr^{-1}$,
showed that a forming star accreting at such rates reaches the zero-age-main-sequence (ZAMS) at a mass of $8-10\,M_\odot$.
Values above this limit can only be reached on the MS, so that in this case no pre-MS massive stars exist,
and thus \hii\ regions can form only around stars that are already on the MS.
\cite{behrend2001} computed models with a time-dependent accretion rate,
starting from the same $\sim10^{-5}M_\odot\rm\,yr^{-1}$ when the star is still in the low-mass range,
but reaching values above $10^{-3}M_\odot\rm\,yr^{-1}$ when the star becomes massive.
With this rate, they were able to reach $\simeq20\,M_\odot$ before the ZAMS.
Moreover, on the Hertzsprung-Russell (HR) diagram, the evolutionary track of a star accreting at such a rate remains relatively close to the ZAMS:
as the stellar mass grows, both luminosity and effective temperature increase.
In this case one expects the radiative feedback of a pre-MS star to be similar to the one of a ZAMS star having the same mass,
allowing \hii\ regions to form around pre-MS stars already.

\cite{hosokawa2009} and \cite{hosokawa2010} computed models with constant accretion rates of $10^{-3}M_\odot\rm\,yr^{-1}$,
and reached the ZAMS at an even higher mass of $30\,M_\odot$.
However, in the models of \cite{hosokawa2009} and \cite{hosokawa2010}, the high accretion rate in the low- and intermediate-mass range
leads to a rapid swelling of the accreting star when its mass is $\sim10\,M_\odot$, and during which the stellar radius reaches $\sim100\,R_\odot$.
When such a swelling occurs, the stellar surface cools down and one expects the ionizing flux to be decreased significantly during this stage.
Using these stellar models, \cite{klassen2012b} deduced the evolution of the size of the corresponding \hii\ region,
and compared it to purely ZAMS models.
They obtained \hii\ regions of significant sizes already in the intermediate-mass range, with both ZAMS and pre-MS models.
However, while the size of the \hii\ region was growing monotonically when they used a ZAMS model,
the results were more complex with the pre-MS models of \cite{hosokawa2009}:
in this case, the swelling of the star reduces the effective temperature,
so that the \hii\ region disappears completely in a few thousand years.
Then, as the star contracts again towards the ZAMS, the effective temperature increases again
and the \hii\ region recovers its pre-swelling size in a timescale of the same order.

The models of \cite{hosokawa2009} are based on the assumption of constant accretion rates.
But all the hydrodynamic simulations for the accretion flow that were computed in three dimensions
show that the accretion rate on forming stars is highly variable, on the same timescale of several thousand years,
due to the formation of spiral arms in the accretion discs
(e.g.~\citealt{banerjee2006,peters2010a,peters2010c,peters2011,kuiper2011,seifried2011,girichidis2012a,klassen2012a}).
Thus we expect the combined effects of time-dependent accretion rates and pre-MS stellar evolution
to be much more complex than those of stellar evolution at constant rate only.
Taking these effects into account requires to compute pre-MS stellar models using the time-dependent accretion rates
obtained from hydrodynamic simulations, instead of constant accretion rates.

Moreover, \cite{haemmerle2016a} showed that the pre-MS evolution of an accreting star
depends sensitively on the choice of the initial model.
In particular, they obtained that for the typical rate $10^{-3}\,M_\odot\rm\,yr^{-1}$ the rapid swelling of the star
described by \cite{hosokawa2009} and \cite{hosokawa2010} does not occur for any initial configuration\footnote{\ 
   We notice that \cite{hosokawa2010} described the effect of changing the initial conditions (see their Sect.~3.2.2).
   But despite the various magnitudes they obtained for the swelling, they concluded that the swelling
   is a strong feature of pre-MS evolution at high accretion rates.
   We notice also that the effect of the initial conditions on the pre-MS evolution at low rates has been studied by \cite{hartmann1997}.}.
Thus one expects the evolution of the ionizing radiation field of accreting stars, and then the properties of the \hii\ regions around them,
to depend significantly on the initial conditions.

In the present work, we explore these various effects.
In Sect.~\ref{sec-st}, we describe the physical inputs and the results of the stellar models;
in Sect.~\ref{sec-hii}, we compute the evolution of the size of the \hii\ regions from the various stellar feedback we obtained;
these results are discussed in Sect.~\ref{sec-dis} and the conclusions are summarized in Sect.~\ref{sec-ccl}.

\section{Stellar models}
\label{sec-st}

\subsection{Accretion histories}
\label{sec-st-ac}

We take accretion histories from a simulation of massive star formation that includes feedback by both ionizing
radiation and by heating of the gas through absorption of non-ionizing radiation by dust
\citep{peters2010a,peters2010b,peters2010c}. We use the instantaneous mass and accretion rate
of seven stars from Run~B of \cite{peters2010a} as input for our pre-MS evolution models.
Three of these stars have reached masses around $20\,M_\odot$
and have produced \hii\ regions in the simulation. Four stars are in the intermediate-mass range and have not
created an \hii\ region yet. The main parameters of these accretion histories are the final mass $M_\mathrm{f}$,
the duration of the accretion period $d_\mathrm{acc}$ and the mean accretion rate $\dot{M}_\mathrm{mean}$.
We summarize these parameters in Table~\ref{tab:acc}.

In these simulations, the UV luminosity is computed with the assumption that the star is on the ZAMS at all
times, so the UV output in these simulations is systematically overestimated. Also, because the UV luminosity grows
monotonically with mass, there is no flickering in the simulations caused by stellar evolution. \cite{klassen2012b}
have investigated the differences in UV output and \hii\ region radius with an improved stellar evolution subgrid model
and detailed stellar evolution calculations and found major differences (their Figure 2). Therefore, the more detailed
stellar evolution calculations presented here would modify the UV output in the original simulations by \citet{peters2010a}.

\begin{table}
\caption{Main parameters of the accretion histories}
\label{tab:acc}
\centering
\begin{tabular}{c c c c}
\hline
\hline
number & $M_\mathrm{f}$ ($M_\odot$) & $d_\mathrm{acc}$ (yr) & $\dot{M}_\mathrm{mean}$ ($M_\odot$ yr$^{-1}$)\\
\hline
1 &  23.4 &      $9.36 \times 10^{4}$&       $2.50 \times 10^{-4}$\\
2 &  7.05 &      $5.22 \times 10^{4}$&       $1.35 \times 10^{-4}$\\
3 &  7.14 &      $9.83 \times 10^{4}$&       $7.27 \times 10^{-5}$\\
4 &  19.5 &      $1.05 \times 10^{5}$&       $1.86 \times 10^{-4}$\\
5 &  21.2 &      $1.22 \times 10^{5}$&       $1.74 \times 10^{-4}$\\
6 &  6.32 &      $1.17 \times 10^{5}$&       $5.41 \times 10^{-5}$\\
7 &  8.06 &      $1.12 \times 10^{5}$&       $7.19 \times 10^{-5}$\\
\hline
\end{tabular}
\end{table}

\subsection{Stellar evolution code}
\label{sec-st-code}

We compute stellar evolution models with the Geneva Stellar Evolution code,
using the seven accretion histories described in Sect.~\ref{sec-st-ac}.
The treatment of accretion in the code has been recently improved (\citealt{haemmerle2016a}),
and a full description of the new treatment is available in \cite{haemmerle2014}.
Notice that a more general description of the code is available in \cite{eggenberger2008}.
Here we only recall briefly the main ideas.

The code is one-dimensional and hydrostatic (i.e. we model only the hydrostatic core, under the accretion shock).
At each timestep $dt$, from $t$ to $t'=t+dt$, the accretion rate $\dot M=dM/dt$ is fixed externally,
by hand or using a pre-defined accretion law, here the accretion histories of Sect.~\ref{sec-st-ac}.
The model $t'$ is obtained from the model $t$ by:
\begin{itemize}
\item computing the new mass $M'=M+dM=M+\dot M\,dt$~;
\item defining the thermal properties of the accreted material~;
\item solving the four equations of stellar structure for $(t',M')$ with the Henyey method.
\end{itemize}

For the thermal properties of the accreted material, we use the assumption of \textit{cold disc accretion}
\citep{palla1992,hosokawa2010},
stating that the entropy of the material accreted during $dt$ is the same as the one of the stellar surface in the model $t$.
This assumption corresponds to the case of an accretion flow having a disc geometry,
in which any entropy excess can be radiated efficiently in the polar directions before it is advected in the stellar interior.
This is the lower limit for the accretion of entropy, the upper limit being the \textit{spherical accretion} (e.g.~\citealt{hosokawa2009}),
in which all the entropy of the accretion flow is advected in the stellar interior.

During the main accretion phase, the stellar timestep $dt$ is much shorter
than the timestep $d\tilde{t}$ of the hydrodynamic simulation.
Thus, in order to conserve the $(t,M)$ relation, we use a stepwise fit for the accretion rate: $\dot M(t):=\dot M(\tilde{t})$,
where $\tilde{t}$ is the time of the hydrodynamic simulation for which $\tilde{t}-d\tilde{t}<t\leq\tilde{t}$.

\subsection{Initial models}
\label{sec-st-ini}

For numerical reasons, we start the computations at the point when the stellar mass is already $2\,M_\odot$.
In the simplified case of a constant accretion rate of $\dot M=10^{-3}\,M_\odot\rm\,yr^{-1}$,
\cite{haemmerle2016a} showed that the evolution of the star during the accretion phase
depends sensitively on the internal structure of the initial model.
In the present work, we used two different initial models in order to check how this choice changes the stellar evolution
during the accretion phase with the more realistic time-dependent accretion rates considered here.

\begin{figure}\includegraphics[width=0.49\textwidth]{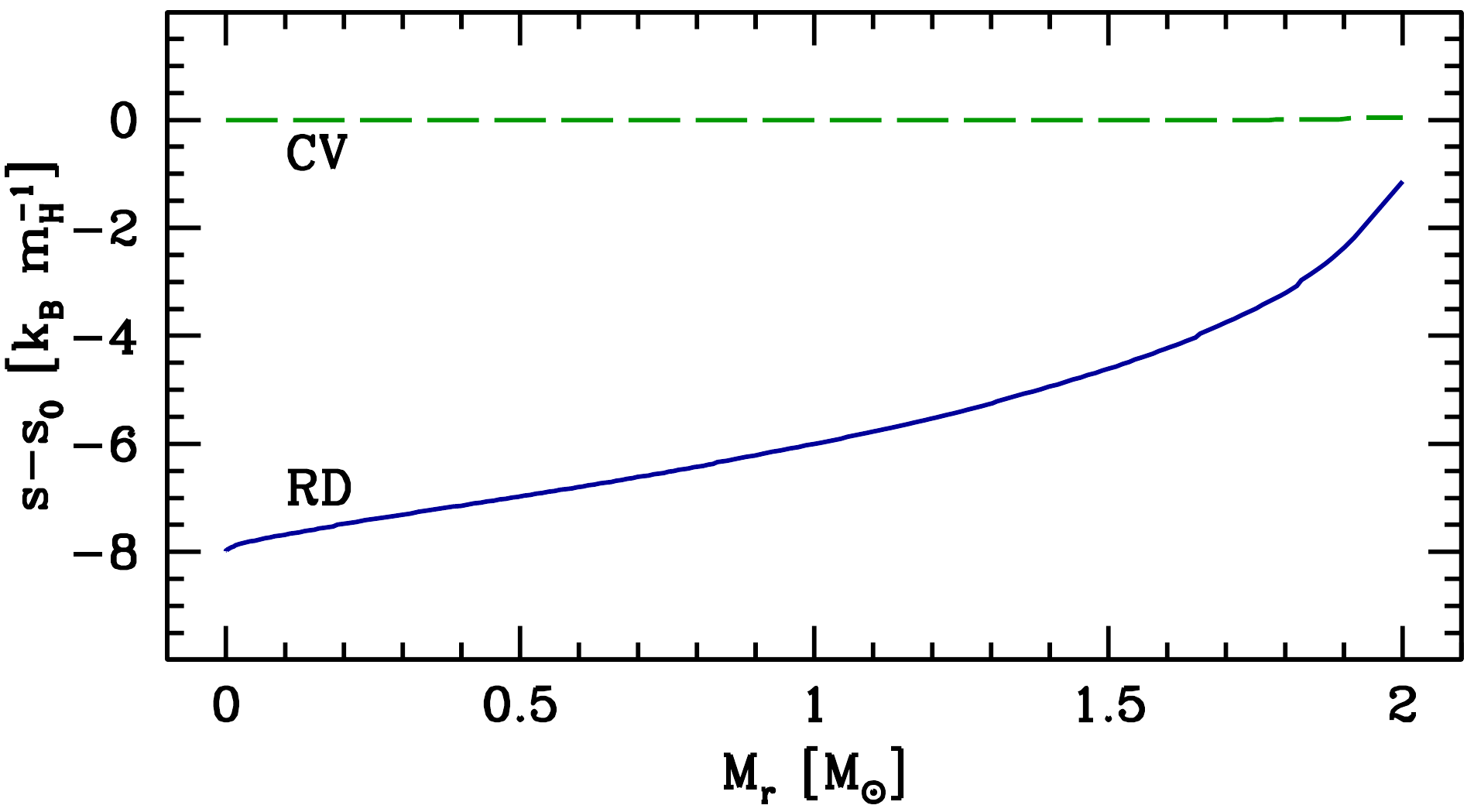}
\caption{Entropy profiles of the two initial models described in Sect.~\ref{sec-st-ini}.
The constant $s_0$ is the (central) entropy of the CV model.}
\label{fig-ini}\end{figure}

The significant difference between these two initial models is the entropy profiles (Fig.~\ref{fig-ini}).
The model CV, which is fully convective, has a flat entropy profile, i.e. convection is adiabatic.
In contrast, the model RD is fully radiative and has thus an entropy gradient $ds/dM_r>0$.
The central entropy in model RD is thus much lower than in model CV.
As a consequence, the model RD is more compact than the model CV.
Indeed, using the homology relations for central pressure and temperature
($P_c\sim M^2/R^4$ and $T_c\sim M/R$), the entropy reads
\begin{equation}
s_c\sim-\ln{P_c\over T_c^{5/2}}\sim\ln(MR^3)
\label{eq-s1}\end{equation}
so that, for a given mass $M$, the lower is the central entropy, the smaller is the radius.
While the model CV has a radius of $20.4\,R_\odot$, the radius of the model RD is $2.9\,R_\odot$ only,
i.e. the radius of RD is smaller by one order of magnitude than the radius of CV.
In terms of surface properties, this difference leads to a much higher luminosity in the CV case
than in the RD case ($L\propto R^2T_{\rm eff}^4$).
In the CV case, we have
\begin{equation}
L=123\,L_\odot		\qquad	T_{\rm eff}=4270\,K
\label{eq-ltcv}\end{equation}
while in the RD case
\begin{equation}
L=9.7\,L_\odot		\qquad	T_{\rm eff}=5960\,K
\label{eq-ltrd}\end{equation}

Moreover, the low central entropy in the model RD corresponds also to a high central temperature.
Indeed, the homology relations give also
\begin{equation}
s_c\sim\ln{M^2\over T_c^{3/2}}
\label{eq-s2}\end{equation}
In the model CV, we have $T_c=7.7\times10^5\,K$, while in RD we have $T_c=9.3\times10^6\,K$.
Due to this high temperature, the opacity is low in RD and the radiative transport is efficient enough for the total flux:
this is the reason why the model RD is not convective.

Another difference between CV and RD is the deuterium abundance.
In the model CV, the value of $T_c$ is lower than the value for significant D-burning,
and we can take the usual ISM mass fraction $X_2=5\times10^{-5}$
(\citealt{bernasconi1996a,norberg2000,behrend2001,haemmerle2013,haemmerle2016a}).
But in contrast to the CV model, the value of $T_c$ in the RD model is much higher than the value for significant D-burning,
and we expect all the deuterium to be destroyed in the previous stages that lead to such an "initial" model.
For this reason, we take $X_2=0$ in the initial model RD.
However, in the accreted material, we keep the value $X_2=5\times10^{-5}$ in both cases.

\quad

We stress that the models considered here as "initial" have a relatively high mass,
and that their properties reflect an earlier evolution which is not treated in the present work.
In the models of \cite{hosokawa2009} and \cite{hosokawa2010},
computed respectively with the assumptions of spherical and disc accretion,
once the stellar mass reaches $2\,M_\odot$, the radius is $\simeq30\,R_\odot$ in the spherical case,
and $\simeq3-4\,R_\odot$ in the disc case.
If we compare these two values with the radii of our CV and RD initial models,
we see that the CV model corresponds roughly to the spherical case and the RD model to the disc case\footnote{\ 
   We notice however that the model of \cite{hosokawa2009} with spherical accretion is fully radiative at $2\,M_\odot$,
   while our CV model is fully convective, despite it has a radius of the same order.}.
Thus we consider the CV model as the case where accretion is spherical until $\simeq2\,M_\odot$
and then proceeds through a disc,
while the RD model corresponds to the case where accretion proceeds through a disc already since $M<<2\,M_\odot$.

\subsection{Results}
\label{sec-st-res}

\subsubsection{Accretion history 1}
\label{sec-st-res-1}

\begin{figure}\includegraphics[width=0.45\textwidth]{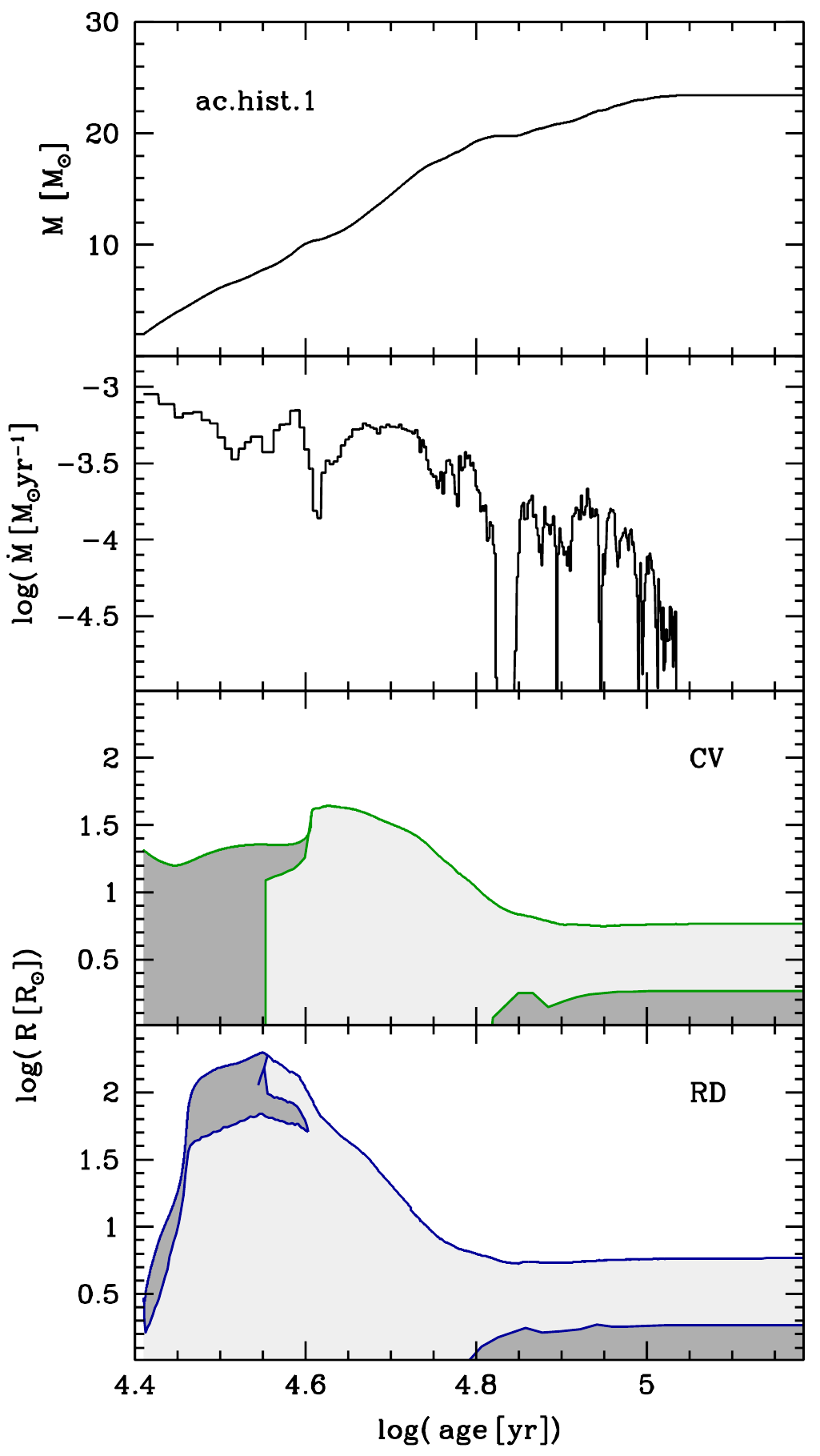}
\caption{Accretion history 1 (mass and accretion rate) and internal structure of the stellar models obtained from it,
for the two initial models described in Sect.~\ref{sec-st-ini}.
On the panels of the internal structure, the upper curve is the stellar radius,
the convective zones are indicated by dark gray areas and the radiative zones by light gray areas.}
\label{fig-st1}\end{figure}

We first consider the evolution of the central star (accretion history 1).
The evolution of the internal structure for both initial models CV and RD is shown on Fig.~\ref{fig-st1},
together with the evolution of the mass and the accretion rate.
In this case, the mass $M=2\,M_\odot$ corresponding to our initial stellar models
is reached at an age\footnote{\ 
   In all the stellar models, we consider the age since the beginning of the hydrodynamic simulation.}
of 25\,700 years.
As we see on Fig.~\ref{fig-st1}, the accretion rate at this "initial" age is relatively high,
with a value $\dot M=9\times10^{-4}\,M_\odot\rm\,yr^{-1}$.
Then it decreases progressively, with several oscillations,
but it remains in the range $[10^{-4},10^{-3}]\,M_\odot\rm\,yr^{-1}$ until an age of 65\,000 years ($\log\rm age\,[yr]=4.81$).
Then it oscillates between 0 and $\sim2\times10^{-4}\,M_\odot\rm\,yr^{-1}$,
vanishing eventually at an age of 110\,000 years ($\log\rm age\,[yr]=5.04$),
leaving the star with its final mass, $M=23.4\,M_\odot$.

In the case CV, the evolution starts with a fully convective structure.
As mentioned in Sect.~\ref{sec-st-ini}, the central temperature is too low for D-burning,
and the star takes its energy from gravitational contraction, loosing entropy ($\epsilon_{\rm grav}=-T\,ds/dt$)
and decreasing its radius (Eq.~\ref{eq-s1}).
After $\sim1000$ years ($M\simeq3\,M_\odot$),
the central temperature reaches $1.5\times10^6\,K$ and D-burning becomes significant.
The thermostatic effect of D-burning slows down the increase of $T_c$,
and the radius starts to increase as the mass grows ($T_c\sim M/R$).
The mass fraction of deuterium decreases with time in the whole star,
until it becomes too small for deuterium to supply through its burning
the energy needed by the star: the central temperature can increase again ($T_c\rightarrow2.5\times10^6\,K$)
and the radius stops growing ($R\simeq22.7\,R_\odot$).
Due to the increase of $T_c$, the opacity decreases in the centre
and at an age of 35\,700 years ($M=7.9\,M_\odot$) a radiative core appears.
The flux emerging from this hot radiative core with low opacity
is too strong for the convective envelope, which is still cold and has a high opacity.
Thus the envelope absorbs a large fraction ($\sim50-80\,\%$) of this flux ($dL_r/dM_r<0$),
its entropy increases ($dL_r/dM_r\simeq-T\,ds/dt$), and the stellar radius too.
The accreted deuterium burns in the convective envelope,
and the entropy released by this shell-burning enhances the swelling\footnote{\ 
   The importance of the contribution from deuterium shell-burning in the evolution is studied in Appendix~\ref{sec-deut}.}.
As the temperature increases in the whole star, the radiative core grows in mass, and the luminosity peak
between the deep regions where $dL_r/dM_r>0$ and the external regions where $dL_r/dM_r<0$
moves outwards (\textit{luminosity wave}, \citealt{larson1972}).
Eventually, the convective envelope disappears and the whole star becomes radiative.
The flux coming from the central regions is no longer absorbed by the envelope,
all the layers of the star are loosing entropy, and the swelling ends:
after the radius reached a maximum of $R_{\rm max}=44\,R_\odot$ (at an age of 42 300 years and a mass of $10.7\,M_\odot$),
it starts to decrease again.
The fully radiative star contracts towards the ZAMS and the central temperature increases from a few $10^6\,K$ to $25\times10^6\,K$.
At this temperature, H-burning becomes significant in the centre, and the energy liberated produces a convective core.
The contraction stops, the radius reaches a minimum of $5.6\,R_\odot$ and the star enters the MS.
The star reaches its final mass of $23.4\,M_\odot$ nearly at this stage: the end of accretion and the ZAMS coincides in this case.

In contrast to the CV case, the evolution in the RD case starts with a fully radiative structure, without deuterium.
However, as soon as the initial model accretes mass, the deuterium contained in this accreted material can burn:
indeed, in this hot initial model, $T=1.7\times10^6\,K$ already at $M_r/M=97\%$.
The energy liberated by this D-burning produces a convective envelope, as it is visible on the lower panel of Fig.~\ref{fig-st1}.
Thus in the RD case, we start the evolution with the luminosity wave and the swelling phase:
a significant fraction of the flux emerging from the hot radiative core is absorbed in the envelope.
But in this case, due to the low surface luminosity (Eq.~\ref{eq-ltrd}),
the envelope radiates this entropy with a much lower efficiency than in the CV case.
As a consequence, the swelling is much stronger: the radius soon exceeds $100\,R_\odot$ (at an age of 29 300 years),
and reaches its maximum value of $198\,R_\odot$ at 35 400 years.
Thus we see that the swelling does not occur at the same age in the RD case than in the CV case.
Then the star contracts again.
In the beginning of the contraction, an intermediate convective zone survives near the surface, due to the remaining D-burning.
But the star becomes rapidly fully radiative, and contracts towards the ZAMS,
which is reached nearly at the end of accretion, as in the CV case.

\begin{figure}\includegraphics[width=0.49\textwidth]{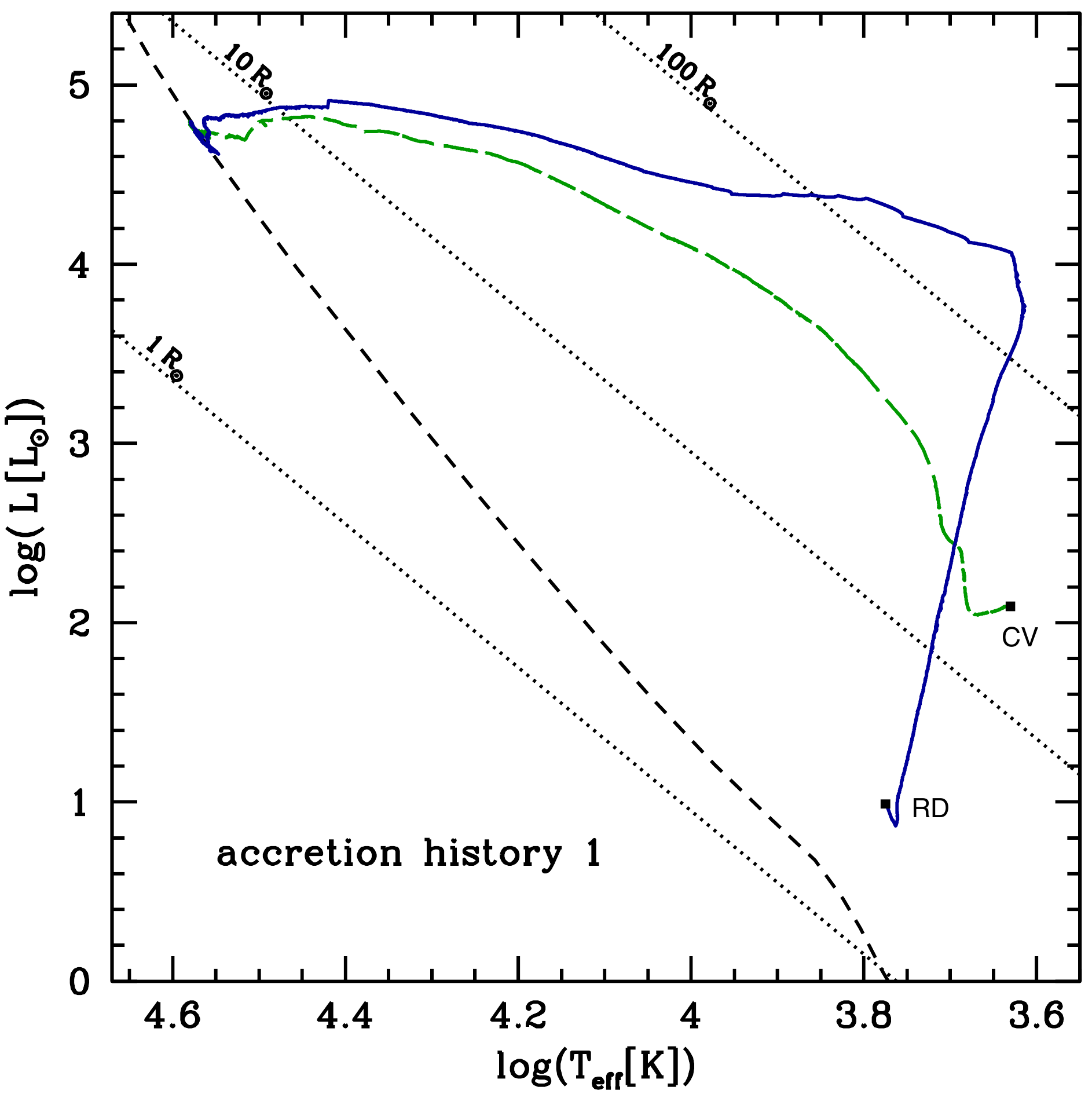}
\caption{Evolutionary tracks obtained from the accretion history 1,
for the two initial models described in Sect.~\ref{sec-st-ini}.
The black dashed curve is the ZAMS \citep{ekstroem2012},
and the black dotted straight lines are iso-radius of 1, 10 and $100\,R_\odot$.}
\label{fig-hr1}\end{figure}

The evolutionary tracks of these two models are shown on Fig.~\ref{fig-hr1}.
As we see, with its relatively weak swelling, the model CV follows a track that is qualitatively similar
to the birthlines at low rates (see e.g.~\citealt{palla1992}), moving leftwards towards the ZAMS (i.e. $T_{\rm eff}$ increases monotonically).
In particular, the luminosity remains under $\sim10^4\,L_\odot$ until the post-swelling contraction,
when the effective temperature increased above $10\,000\,K$.
In contrast, the model RD, with its strong swelling,
starts its evolution along a nearly straight line towards the luminous and red regions of the HR diagram (i.e. $T_{\rm eff}$ decreases).
At the maximum of the swelling, the star reaches a luminosity of $\sim10^4\,L_\odot$
with an effective temperature as low as $4000\,K$.
Then, as the star contracts, it moves leftwards towards the ZAMS (i.e. $T_{\rm eff}$ increases).

In Appendix~\ref{sec-rc} we show that it is indeed the size of the initial radiative core, and not the size
of the total initial stellar radius, that determines the magnitude of the swelling.

\subsubsection{Accretion histories 2 to 7}
\label{sec-st-res-27}

\begin{figure}\includegraphics[width=0.49\textwidth]{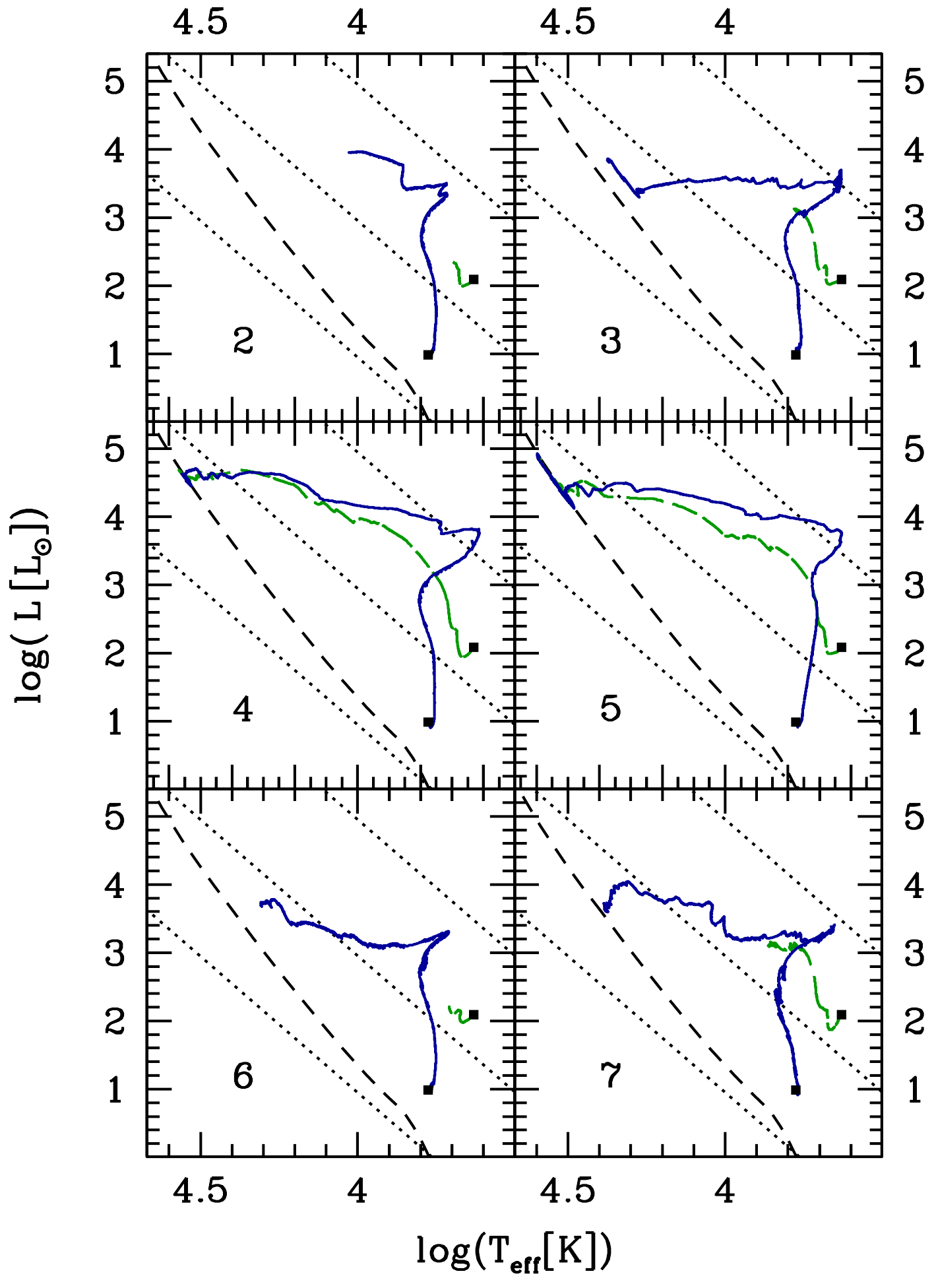}
\caption{Same as Fig.~\ref{fig-hr1}, but for the accretion histories 2 to 7.}
\label{fig-hr27}\end{figure}

The evolution of the internal structure for accretion histories 2 to 7 is qualitatively similar to the case of accretion history 1.
The evolutionary tracks are shown on Fig.~\ref{fig-hr27}.

For accretion histories 2, 3, 6 and 7, the star is still in the intermediate-mass range ($\simeq7\,M_\odot$)
at the end of the simulation (due to a late core-formation in the cases 2 and 3, and to a low accretion rate in the cases 6 and 7).
As a consequence, in the cases 2 and 6, the CV model did not enter yet the swelling phase when we stopped the computation.
In the cases 3 and 7, the computation ends soon after a local maximum of the radius,
which could potentially be the maximum, depending on the subsequent accretion episodes.
For histories 4 and 5, the core forms at an early age ($\sim30\,000\rm\,yr$)
and the accretion rate is relatively high ($\sim5\times10^{-4}\,M_\odot\rm\,yr^{-1}$),
so that the star can enter the high-mass range ($M\simeq20\,M_\odot$ at the end of the simulation).
In these cases, the star follows the whole pre-MS phase
and reaches the ZAMS just before the end of the simulation (in both CV and RD cases).
In all the six accretion histories, the CV model never reaches radii higher than $50\,R_\odot$,
and the luminosity exceeds $10^4\,L_\odot$ only for $T_{\rm eff}>10\,000\,K$:
the evolutionary tracks remain similar to the case of low $\dot M$, with a star moving leftwards towards the ZAMS.

In the RD case, the star reaches the end of the swelling for all the accretion histories.
As in the case of the accretion history 1, the hot and compact initial model RD starts its evolution with the swelling,
while in the CV case the star is still burning slowly its deuterium in the centre.
As a consequence, the swelling occurs at a younger age in the RD case than in the CV case,
and the maximum radius at the end of the swelling in the RD case is above $50\,R_\odot$ for all the accretion histories,
in contrast to the CV case.
Moreover, the higher is the average accretion rate, the larger is the maximum value of the radius.
For histories 3, 4 and 5, in which $\dot M$ exceeds or approaches $5\times10^{-4}\,M_\odot\rm\,yr^{-1}$,
the maximum radius of the RD model is even above $100\,R_\odot$, as for accretion history 1.
(In histories 2, 6 and 7, we stay at $\dot M\lesssim2-3\times10^{-4}\,M_\odot\rm\,yr^{-1}$, and $R<100\,R_\odot$.)
In each of these six cases, the RD evolutionary track shows a rightwards turn during the swelling,
leading the star to the red.
For histories 4 and 5, the luminosity approaches $10^4\,L_\odot$ with an effective temperature as low as $4000-5000\,K$,
due to this turn, while the CV model reaches such luminosities only when $T_{\rm eff}>10\,000\,K$.
We notice that the models that reach the ZAMS before the end of the simulation in the CV case
(histories 4 and 5) reach also the ZAMS in the RD case, and vice versa.

\section{\hii\ regions}
\label{sec-hii}

\subsection{Ionizing flux}
\label{sec-hii-ion}

In order to evaluate the effect of the stellar feedback on the size of a surrounding \hii\ region,
we first compute the number of ionizing photons per second by integrating the black-body spectrum above the ionizing energy:
\begin{eqnarray}
S^*&=&4\pi R^2\int\limits_{\scriptscriptstyle h\nu>13.6\rm\,eV}{F_\nu\over h\nu}\,d\nu \nonumber\\
&=&{8\pi^2R^2\over c^2h^2}\int\limits_{\scriptscriptstyle h\nu>13.6\rm\,eV}{(h\nu)^2\over e^{h\nu/kT_{\rm eff}}-1}\,d\nu
\label{eq-ion}\end{eqnarray}

\begin{figure}\includegraphics[width=0.45\textwidth]{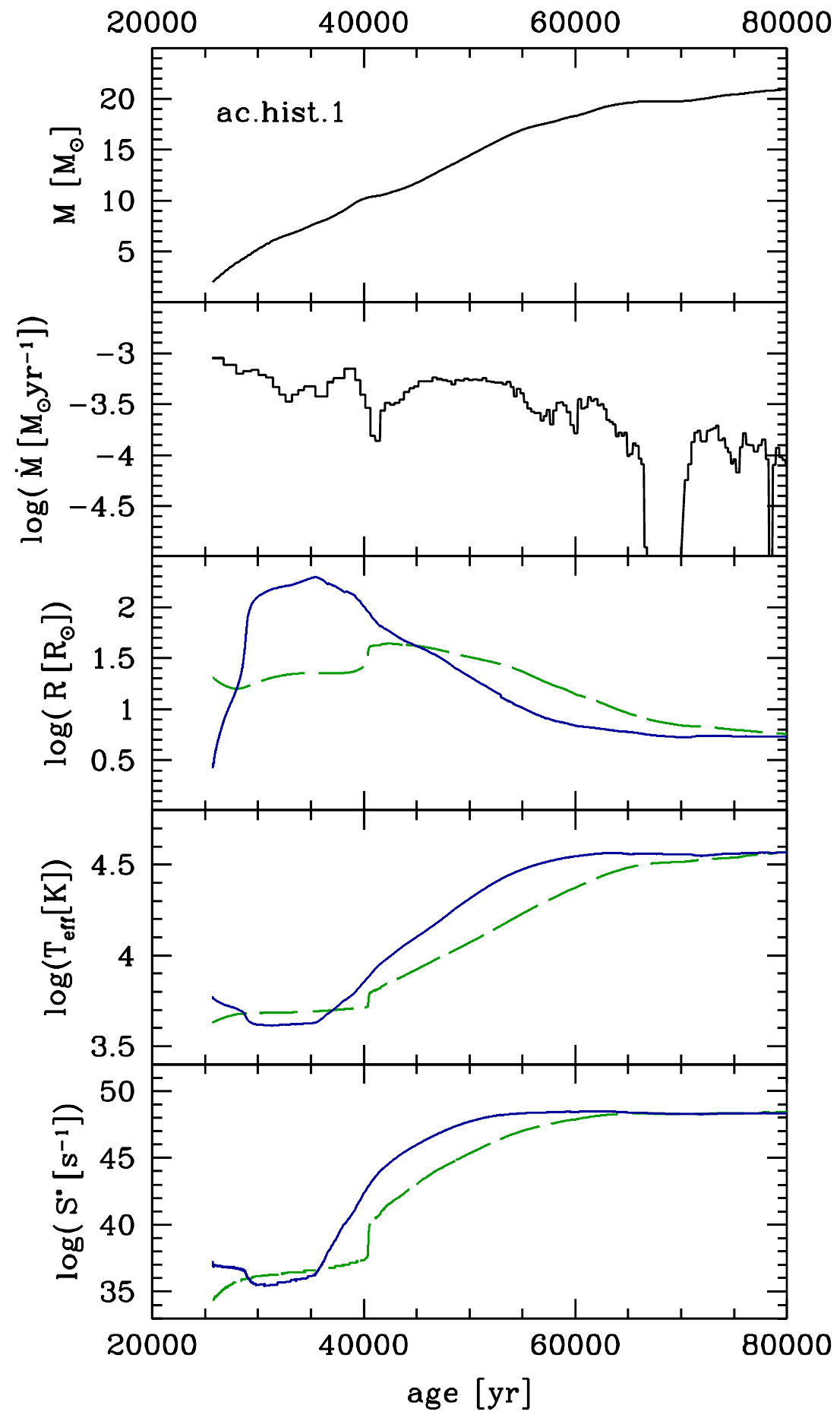}
\caption{Evolution of the mass, the accretion rate, the stellar radius, the effective temperature and the flux of ionizing photons (Eq.~\ref{eq-ion})
for accretion history 1 and the two initial models CV (green dashed line) and RD (solid blue line).}
\label{fig-ion1}\end{figure}

\begin{figure}\includegraphics[width=0.45\textwidth]{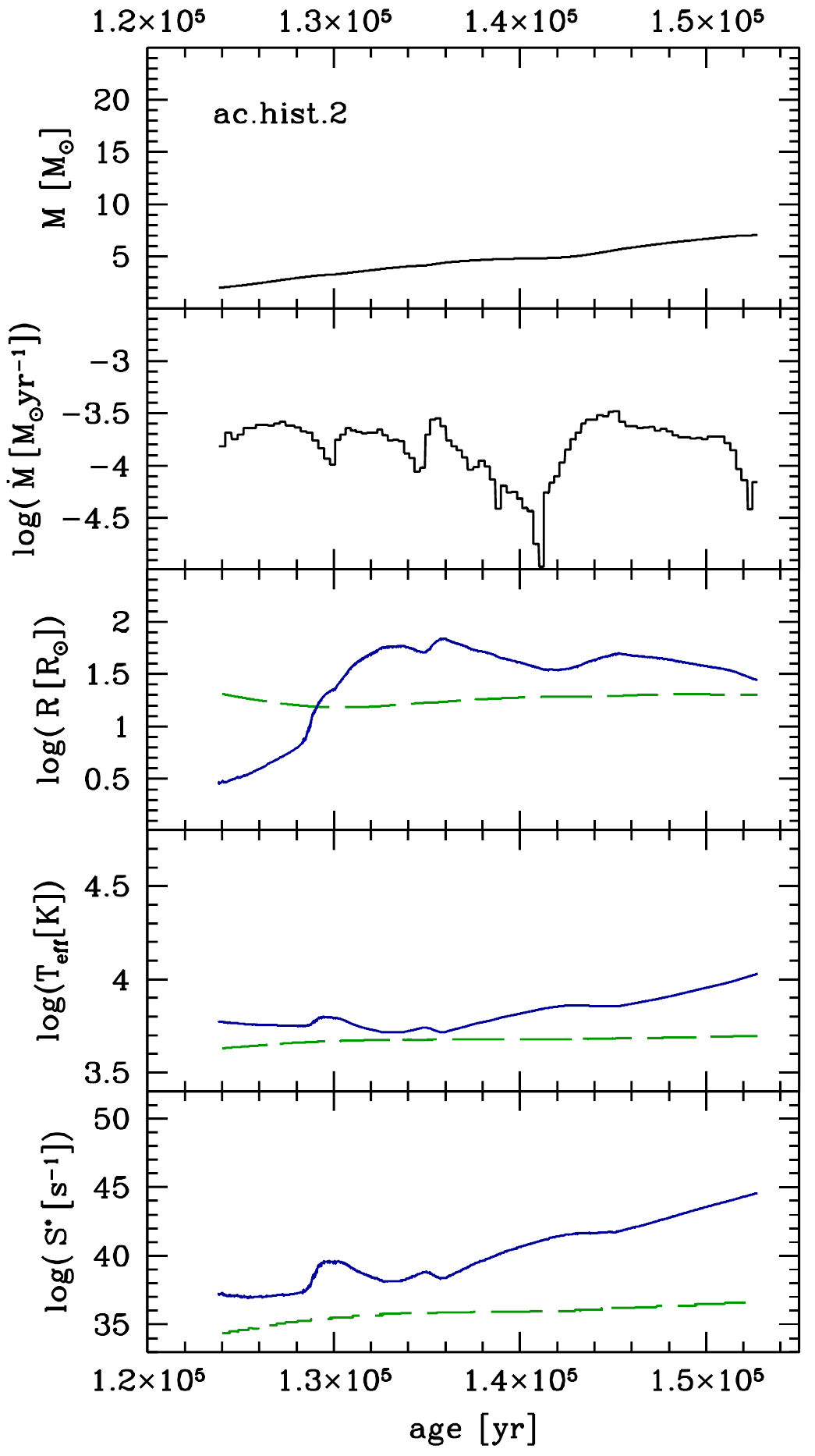}
\caption{Same as Fig.~\ref{fig-ion1} for accretion history 2.}
\label{fig-ion2}\end{figure}

\begin{figure}\includegraphics[width=0.45\textwidth]{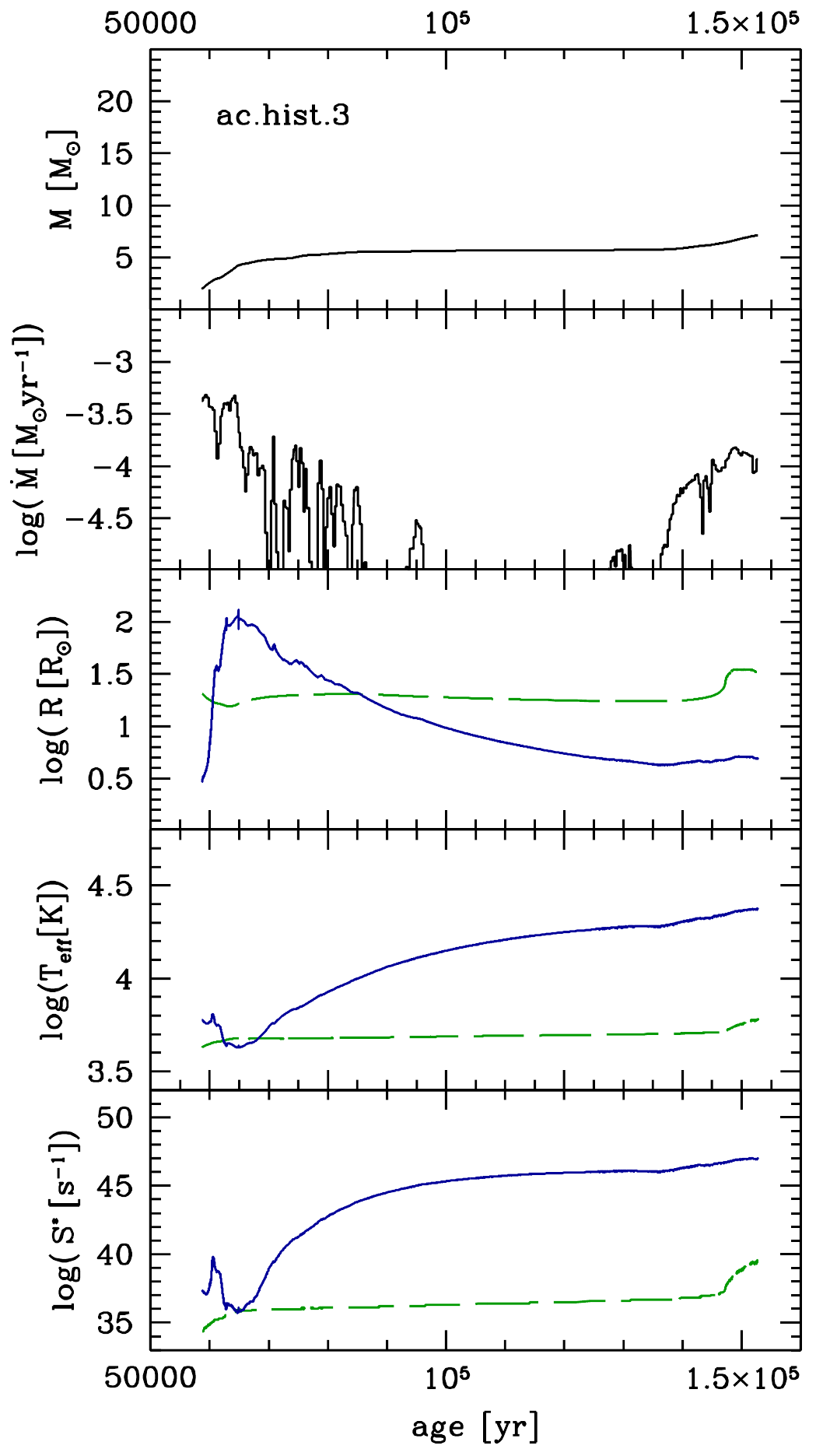}
\caption{Same as Fig.~\ref{fig-ion1} for accretion history 3.}
\label{fig-ion3}\end{figure}

\begin{figure}\includegraphics[width=0.45\textwidth]{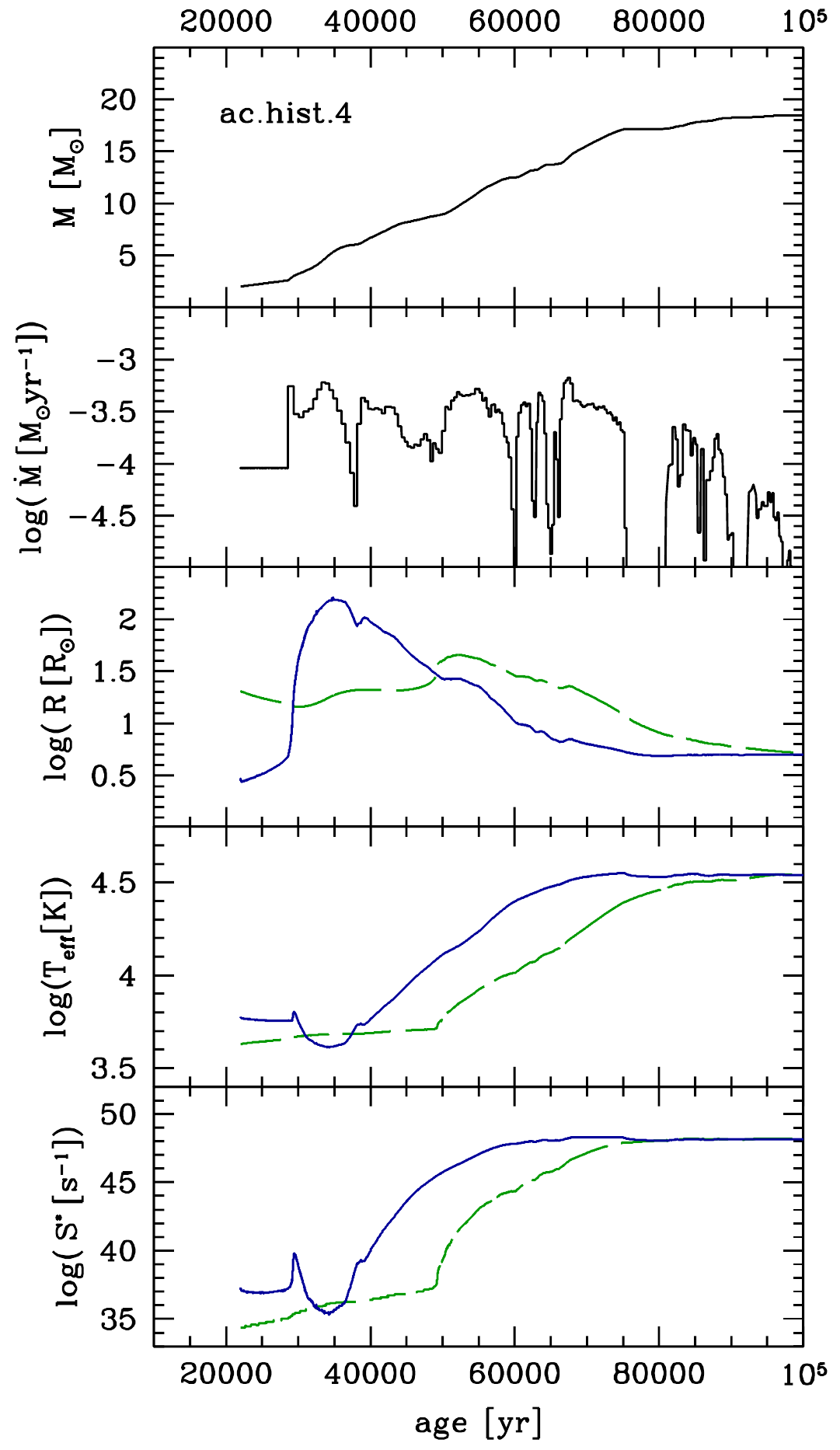}
\caption{Same as Fig.~\ref{fig-ion1} for accretion history 4.}
\label{fig-ion4}\end{figure}

\begin{figure}\includegraphics[width=0.45\textwidth]{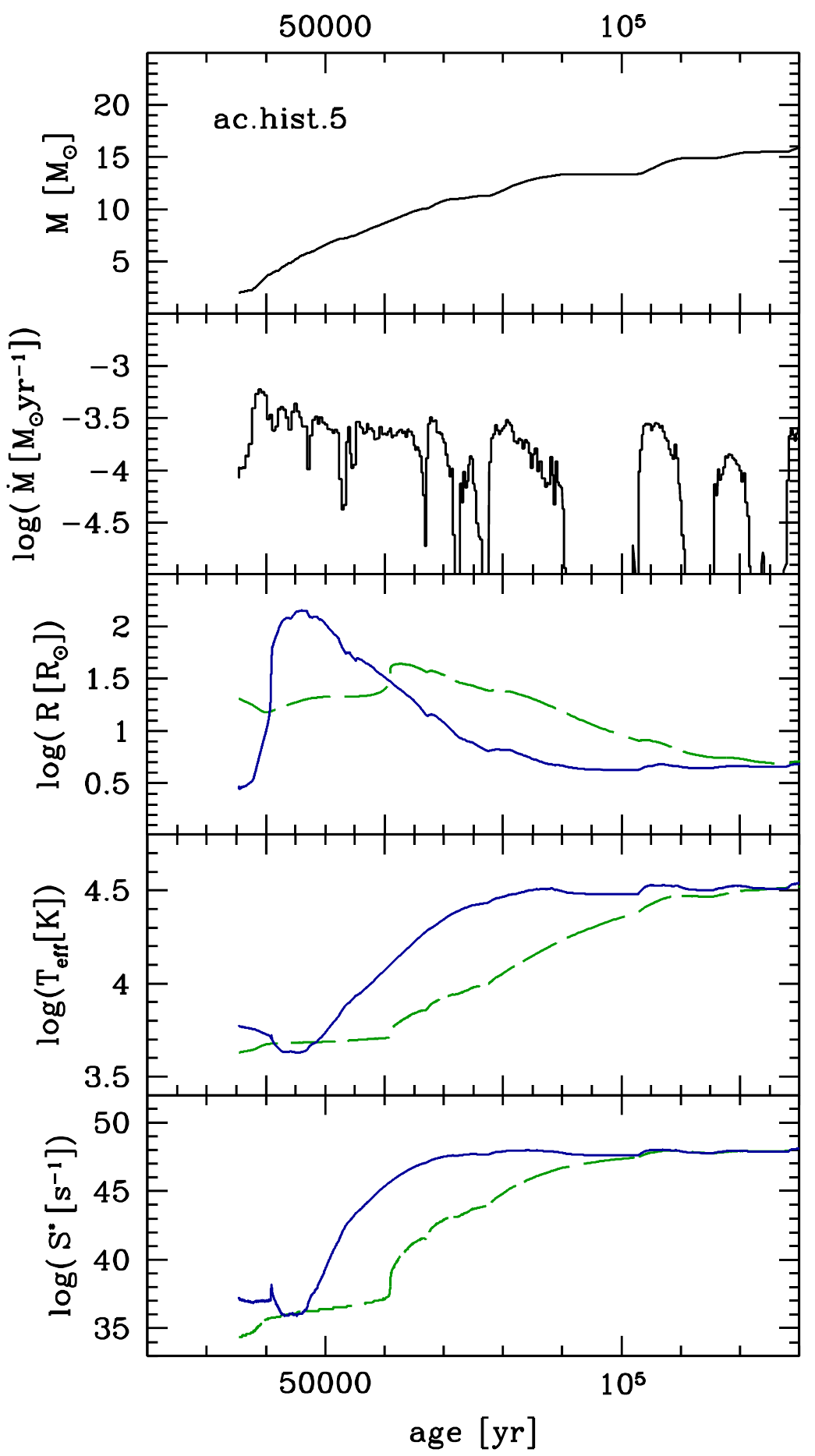}
\caption{Same as Fig.~\ref{fig-ion1} for accretion history 5.}
\label{fig-ion5}\end{figure}

\begin{figure}\includegraphics[width=0.45\textwidth]{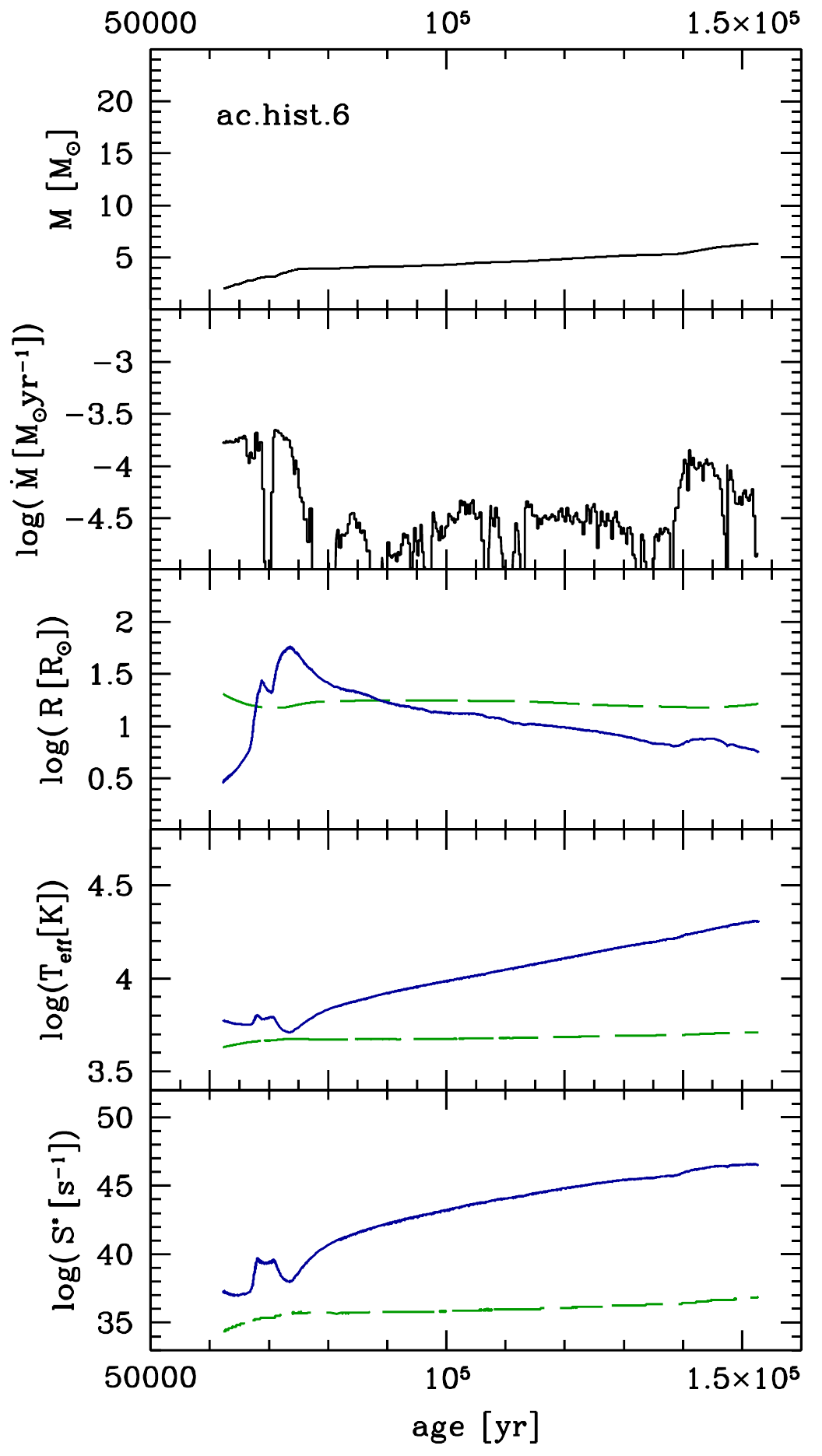}
\caption{Same as Fig.~\ref{fig-ion1} for accretion history 6.}
\label{fig-ion6}\end{figure}

\begin{figure}\includegraphics[width=0.45\textwidth]{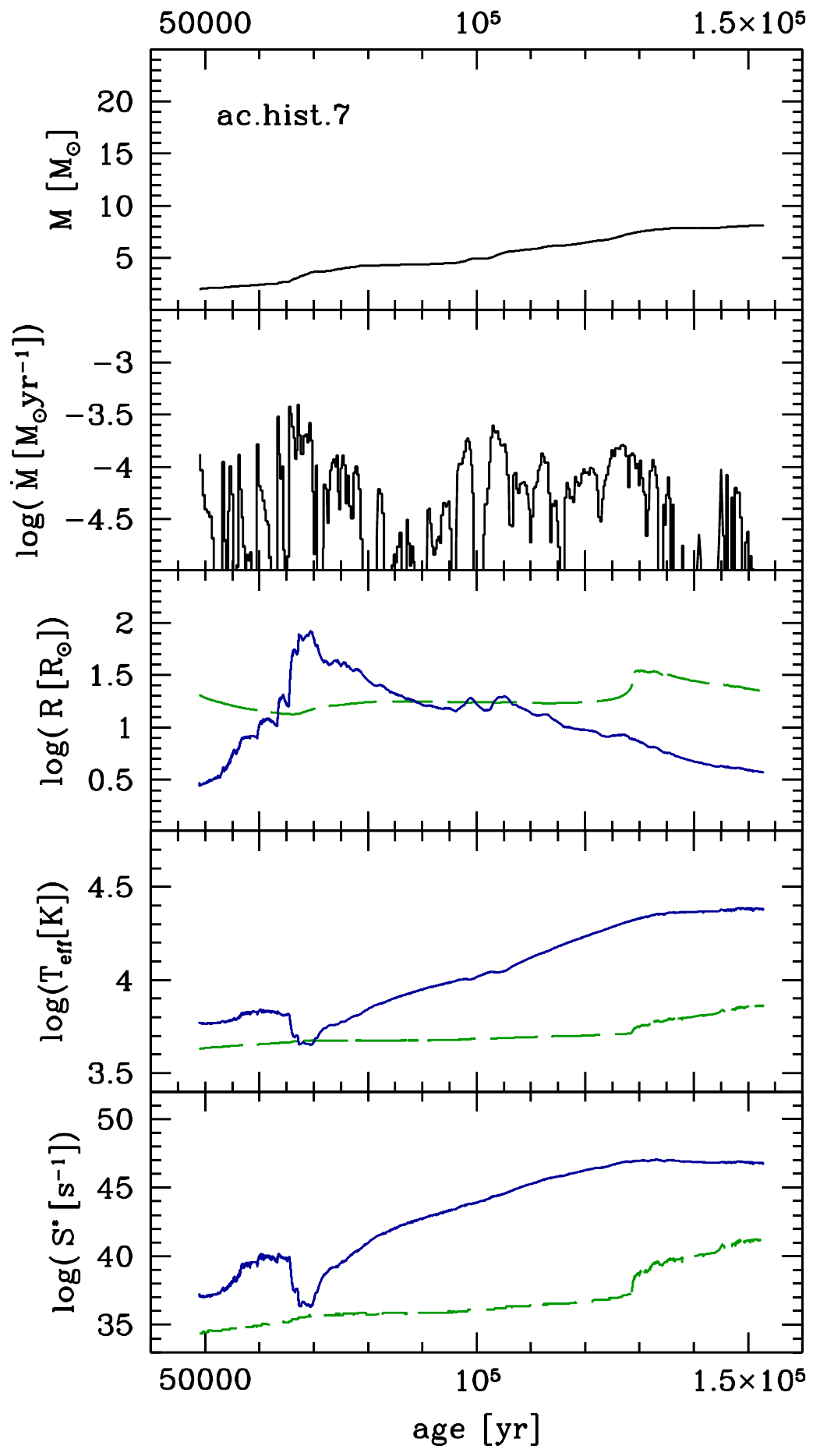}
\caption{Same as Fig.~\ref{fig-ion1} for accretion history 7.}
\label{fig-ion7}\end{figure}

The evolution of $S^*$ for the seven accretion histories and the two initial models is shown on Fig.~\ref{fig-ion1} to \ref{fig-ion7}.
In order to understand the evolution of the ionizing flux,
we plot also the accretion rate, the stellar radius and the effective temperature on the same figures.
All these quantities are shown as a function of the age.
As mentioned in Sect.~\ref{sec-st-res-1} and \ref{sec-st-res-27}, for a given accretion history,
the swelling does not occur at the same age in the CV and RD cases,
so that one has to be careful when comparing the evolution of $S^*$ between both cases.
In particular, for accretion histories 2, 3, 6 and 7, in the CV case
the star does not go through the whole swelling and post-swelling phase before the end of the simulation.

As mentioned in Sect.~\ref{sec-st-res-1} and \ref{sec-st-res-27}, in the CV case
the effective temperature increases monotonically for all the accretion histories considered.
In terms of the ionizing flux, it results in a monotonous increase in $S^*$.
Before the swelling, it remains $S^*\lesssim10^{37}\rm\,s^{-1}$,
but when the swelling occurs, the rapid increase in $T_{\rm eff}$ and $L$ leads to a jump in $S^*$, by several orders of magnitude.
The jump is particularly abrupt for accretion history 1:
when the swelling becomes the stronger, $S^*$ increases by two orders of magnitude in 100 years only.
In all the accretion histories for which the CV model reaches the end of the swelling,
$S^*$ exceeds $10^{40}\rm\,s^{-1}$ at the end of the swelling, except for accretion history 7.
Then, during the post-swelling contraction, it continues to increase more slowly,
by several orders of magnitude in typically 20 000 years, reaching $S^*\simeq10^{48}\rm\,s^{-1}$ on the ZAMS.

In contrast, as mentioned in Sect.~\ref{sec-st-res-1} and \ref{sec-st-res-27},
the effective temperature does not increase monotonically in the RD case,
and thus the behaviour of $S^*$ is much more complex than in the CV case described above.
In the beginning of the simulations, which corresponds to the early swelling phase in the RD case,
the evolution of the ionizing flux is very sensitive to the evolution of the accretion rate.
For accretion history 1, in which the swelling is strong already at the very beginning of the simulation,
the evolution starts with a decreasing $T_{\rm eff}$, and thus $S^*$ decreases slowly.
The swelling is still enhanced, the decrease in $T_{\rm eff}$ accelerates
and $S^*$ falls suddenly by one order of magnitude in less than 1000 years.
Then the swelling slows down, the star starts to contract again, and thus $T_{\rm eff}$ and $S^*$ can increase.
Roughly 5000 years after its sudden fall, $S^*$ reaches again its previous value.
The star contracts towards the ZAMS, which is reached with the same $S^*\simeq10^{48}\rm\,s^{-1}$ as in the CV case.

For the other six accretion histories, the early behaviour of $S^*$ is much richer, showing rapid oscillations:
since the swelling is initially weaker, the evolution of the effective temperature in the early stages depends more sensitively
on the variations of the accretion rate.
In all these six cases, in the beginning of the swelling, a lowering of the accretion rate is responsible for a jump in the ionizing flux:
a sudden fall in $\dot M$ slows down the swelling, the effective temperature increases and thus $S^*$ shows a jump.
For all these histories except the 5, $S^*$ reaches a value as high as $\simeq10^{40}\rm\,s^{-1}$ during this event.
It remains at such a high value during a timescale of $\sim1000$ years (500 years for accretion history 5, 10 000 years for accretion history 7)
Then the reverse effect occurs: as the accretion rate increases again, the swelling is enhanced, the effective temperature falls abruptly,
and the ionizing flux is reduced.
It happens for all the accretion histories from 2 to 7, but it is particularly noticeable for histories 3, 4, 5 and 7,
in which $S^*$ is reduced under its initial value.
This decrease occurs also on a timescale of $\sim1000$ years (1000 years for accretion history 3, 4000 years for accretion history 4).
Then the swelling ends, $T_{\rm eff}$ starts to increase again and $S^*$ grows slowly, as the star contracts towards the ZAMS.
Again, the timescale between the sudden fall and the slow increase in $S^*$ is a few $1000$ years
(3000 years for accretion history 6, 5000 years for accretion history 3).
When the star reaches the ZAMS, its ionizing flux is the same as in the CV case ($S^*\simeq10^{48}\rm\,s^{-1}$, as mentioned above).

\quad

We notice that for the three accretion histories that lead to massive stars (1, 4 and 5),
the main increase in the ionizing flux that occurs during the post-swelling contraction
is delayed by $\sim10,000$ years in model CV compared to model RD.
This is a consequence of the shift in the age at which the swelling occurs, as mentioned above.
It suggests that this shift produces also a similar delay in the expansion of the \hii\ region.

\subsection{Str\"omgren radius}
\label{sec-hii-strom}

We estimate the size of the \hii\ regions produced by the ionizing feedbacks described in the previous section.
We use the Str\"omgren approximation, based on the assumption that the number of ionizing photons emitted per second
equals the number of recombinations within the volume of the \hii\ region (see e.g.~\citealt{dyson1997,spitzer1998}).
The Str\"omgren radius is given by
\begin{equation}
S^*={4\over3}\pi\,R_{\mathrm{S}}^3\,\alpha\,n_{\rm el}^2
\label{eq-rs}\end{equation}
where $\alpha$ is the recombination rate and $n_{\rm el}$ the electron number density.
We take $\alpha=2.59\times10^{-13}\rm\,cm^3\,s^{-1}$, corresponding to a gas temperature of $\sim10^4\,K$.
For the electron number density, we assume a constant value, but we consider two different cases:
$n_{\rm el}=10^4$ and $10^6\rm\,cm^{-3}$, corresponding respectively to ultra-compact (UCHII)
and hyper-compact (HCHII) \hii\ regions (\citealt{kurtz2005}).
With these assumptions, we compute $R_{\mathrm{S}}$ from the ionizing fluxes $S^*$ described in the previous section, using Eq.~(\ref{eq-rs}).
When $R_{\mathrm{S}}<R$, we impose $R_{\mathrm{S}}=R$.

\begin{figure*}\includegraphics[width=0.99\textwidth]{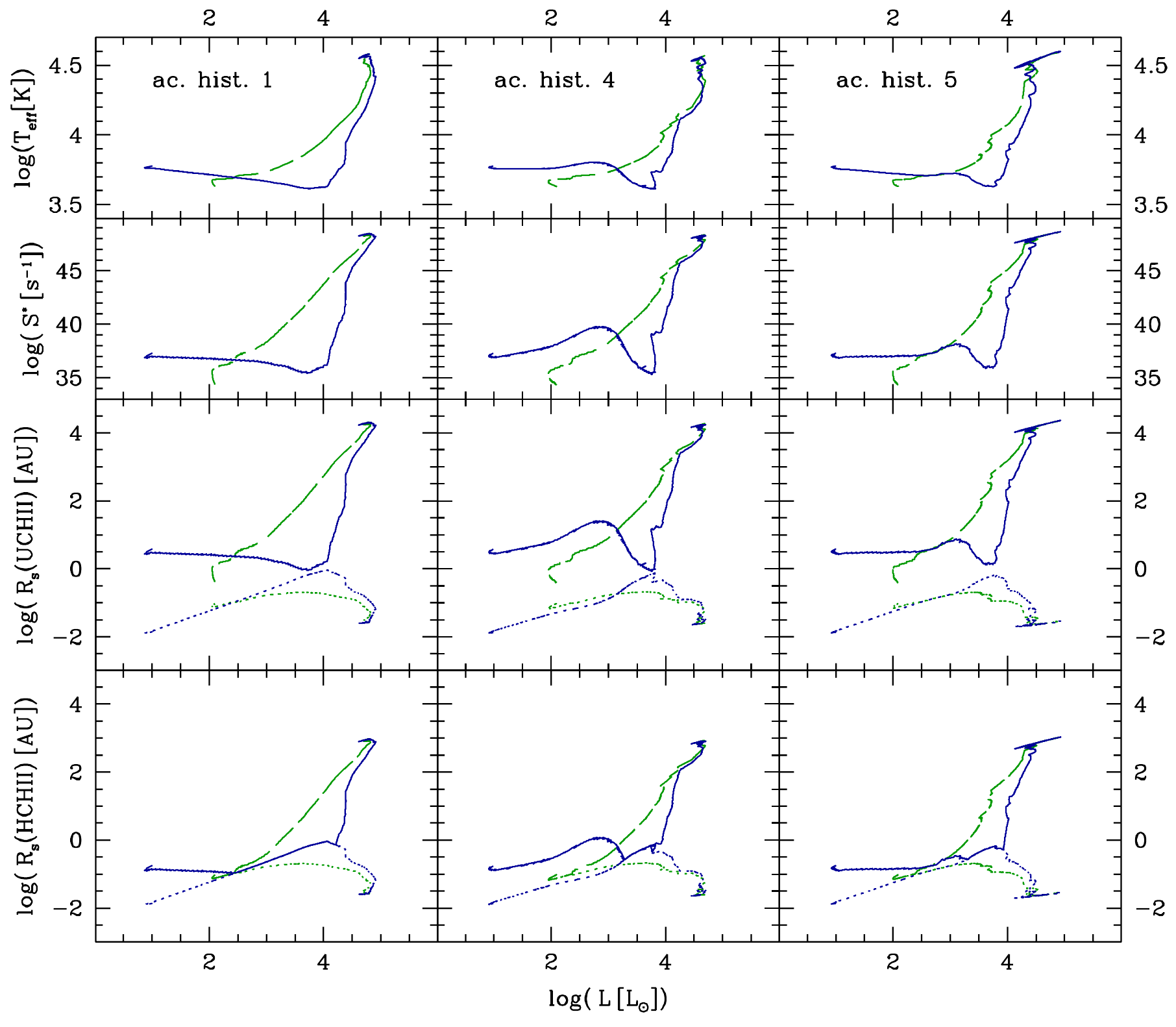}
\caption{Evolution of the effective temperature, the ionizing flux and the Str\"omgren radius
for accretion histories 1, 4 and 5 and for both CV (green) and RD (blue) initial models.
The Str\"omgren radius is computed using $n_{\rm el}=10^4\rm\,cm^{-3}$ in the UCHII case,
and $n_{\rm el}=10^6\rm\,cm^{-3}$ in the HCHII case (\citealt{kurtz2005}).
The dotted curves on the panels of the Str\"omgren radius indicate the stellar radius.}
\label{fig-rs}\end{figure*}

The result is shown on Fig.~\ref{fig-rs} for the three accretion histories leading to masses $\sim20\,M_\odot$.
We used the luminosity for the horizontal axis, in order to compare models with observational quantities.
Eq.~(\ref{eq-rs}) gives $\log R_{\mathrm{S}}={1\over3}\log S^*+\rm cst$,
so that on a logarithmic scale the curve of $R_{\mathrm{S}}$ is simply a re-scaling of the curve of $S^*$.

As we see, for these three accretion histories, the $L-S^*$ and $L-R_{\mathrm{S}}$ relations differ significantly between the CV and RD cases.
In the CV case, due to the early monotonous increase in $T_{\rm eff}$, the ionizing flux becomes rapidly high,
reaching $S^*\gtrsim10^{40}\rm\,s^{-1}$ as soon as $\log(L/L_\odot)\gtrsim3.5$, in all three accretion histories.
At this stage, the Str\"omgren radius $R_{\mathrm{S}}\rm(UCHII)$ is already $\simeq30\rm\,[AU]$.
In contrast, in the RD case, the decrease in $T_{\rm eff}$ during the strong swelling
leads the ionizing flux to values as low as $S^*\sim10^{35}-10^{36}\rm\,s^{-1}$
when the bolometric luminosity is already $\log(L/L_\odot)\simeq3.8$.
For such a low value of $S^*$, we have $R_{\mathrm{S}}\rm(UCHII)\simeq1\rm\,[AU]$ only.
Due to the strong swelling in the RD case, we have at this point $R_{\mathrm{S}}{\rm(UCHII)}\simeq R$,
and hydrogen is ionized only in a relatively thin shell around the star.
For accretion history 1, the thickness of this \hii\ shell remains as low as 0.5\,[AU] during 5000 years.
Then $T_{\rm eff}$ increases in both CV and RD cases, and the models converge to
$R_{\mathrm{S}}\rm(UCHII)\simeq10^4\,[AU]\simeq0.05\,[pc]$ as they approach the ZAMS.

When we increase $n_{\rm el}$ by two orders of magnitude, we decrease $R_{\mathrm{S}}$ by more than one order of magnitude (Eq.~\ref{eq-rs}).
On the lower panel, the Str\"omgren radius is shown for $n_{\rm el}=10^6\rm\,cm^{-3}$ instead of $n_{\rm el}=10^4\rm\,cm^{-3}$.
In this case, for the RD model, at the maximum of the swelling the Str\"omgren radius becomes smaller than the stellar radius
($R\simeq1\rm\,[AU]$), i.e. the \hii\ region disappears completely, "engulfed" in the star.
This events lasts 10\,000 years, until the stellar luminosity reaches $\simeq10^4\,L/L_\odot$.
For the CV model, with its weaker swelling, $R_{\mathrm{S}}\rm(HCHII)$ reached already nearly 100\,[AU] at the same luminosity.
Then the contraction of the star leads to an increase in $T_{\rm eff}$,
and $R_{\mathrm{S}}\rm(HCHII)$ reaches $\simeq1000\rm\,[AU]$ when the star approaches the ZAMS, for both CV and RD models.

\section{Discussion}
\label{sec-dis}

\subsection{Magnitude of the swelling}
\label{sec-dis-sw}

The results of the previous sections show that, for a given accretion history,
the evolution of the star during the pre-MS depends sensitively on the initial model.
In particular, the existence of a significant swelling, that leads to radii $\gtrsim100\,R_\odot$ and brings back the star to the red,
depends on the initial entropy profile.
In each case, the hot, compact and radiative model RD leads to significantly higher radii than the cold, extended and convective model CV.
While the CV tracks remain qualitatively similar to the classical tracks at low $\dot M$, with the star moving leftwards,
the RD tracks show a rightwards turn during the swelling, leading the star to the red, for each of the seven accretion histories.

\begin{figure}\includegraphics[width=0.49\textwidth]{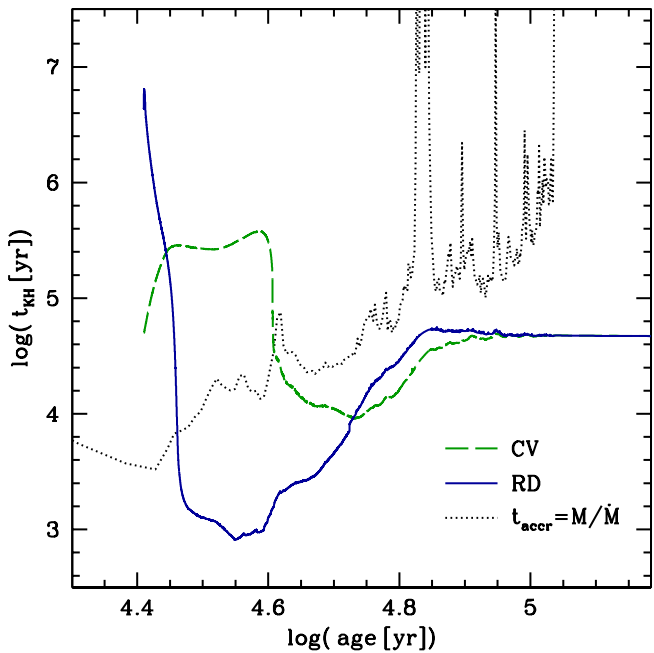}
\caption{Timescales for the same models as on Fig.~\ref{fig-st1}.
The green and blue curves indicate the Kelvin-Helmholtz time,
while the black dotted lines is the accretion time $M/\dot M$, given by the accretion history.}
\label{fig-t}\end{figure}

Since the swelling is produced by the luminosity wave (see Sect.~\ref{sec-st-res-1}),
which corresponds to an internal redistribution of the entropy that restores thermal equilibrium,
we expect the difference in the magnitude of the swelling between the CV and the RD models to take its origin
in the relative efficiency of the various layers of the star to transport and radiate their entropy.
This is illustrated by the timescales for the loss of entropy by radiation (i.e. the Kelvin-Helmholtz time)
\begin{equation}
t_{\rm KH}={GM^2\over RL}
\label{eq-tkh}\end{equation}
and for the mass growth by accretion (i.e. the accretion time)
\begin{equation}
t_{\rm accr}={M\over\dot M}
\label{eq-tac}\end{equation}
These timescales are shown on Fig.~\ref{fig-t} for the accretion history 1, for both CV and RD cases.
(Since $t_{\rm accr}$ depends only on the accretion history, it is the same in both cases.)
As we see, in both cases, the evolution starts with $t_{\rm KH}>>t_{\rm accr}$.
In other words, the star "accretes mass faster than it radiates entropy",
and it has not the time to adjust thermally to the newly accreted material, producing the swelling
(i.e. the increase in $R$ produced by the increase in $M$ dominates over the decrease in $R$ due to the decrease in $s$).
In both cases, the increase in $R$ and $L$ during the swelling reduces $t_{\rm KH}$ by several orders of magnitude,
and brings it back under $t_{\rm accr}$.
Then the loss of entropy by radiation becomes efficient enough for the star to contract again, and the swelling ends.
By comparing $t_{\rm KH}$ at the beginning of the simulation in the CV and RD cases,
we see that it is much longer in the RD case than in the CV case,
by two orders of magnitude, due to the low $R$ and $L$ in the RD case.
As a consequence, the thermal adjustment by surface radiation is less efficient in the RD case than in the CV case,
so that the swelling is stronger, leading to the rightwards turn on the HR diagram, as described above.
It explains also the fact that, the higher is the accretion rate, the stronger is the swelling.
Indeed, a high accretion rate gives a short accretion time (Eq.~\ref{eq-tac}),
so that the higher is $\dot M$, the more difficult it is for the star to adjust thermally to the mass increase.

As demonstrated in App.~\ref{sec-rc}, the global quantities such as $t_{\rm KH}$ do not allow
to understand the effect of a change in the initial configuration on the magnitude of the swelling in all the cases.
A detailed comparison between the initial models would require to consider local quantities
and the internal structures instead of only global quantities.
However, when we compare two extreme cases like the CV and the RD models,
we still expect $t_{\rm KH}$ to be a relevant indicator to express the ability of the star to restore thermal equilibrium,
provided that we consider its value before the swelling.
Once the swelling occurs, the contraction of the stellar interior is highly non-homologous, and $t_{\rm KH}$ is much less relevant.
In particular, we see on Fig.~\ref{fig-t} that the swelling of the RD model does not stop when $t_{\rm KH}$
becomes shorter than in the CV model, and that it continues even several thousand years after $t_{\rm KH}<t_{\rm accr}$.

Finally, we notice that, since the swelling occurs earlier for the RD model than for the CV one,
the mass at which the swelling starts is also lower and the accretion time shorter in the RD case than in the CV case.
Thus one could wonder how this difference influences the results.
In App.~\ref{sec-taccr}, we show that this effect is negligible.

\subsection{Initial model and early accretion geometry}
\label{sec-dis-geo}

As mentioned in Sect.~\ref{sec-st-ini}, our two initial models can be seen as models built by spherical (CV) and disc (RD) accretion.
The significant differences in the evolution of the star and the surrounding \hii\ region that we obtain between these two cases
suggest that a change in the accretion geometry during the early accretion phase only, has a critical impact.
In the case of a purely disc geometry (RD case) we obtain a strong swelling ($R\sim100\,R_\odot$)
for all the accretion histories considered.
In contrast, in the case of an accretion geometry that is spherical in the early stages (i.e. in the low-mass range $M<2\,M_\odot$)
and then switched to a disc geometry for $M>2\,M_\odot$ (CV case),
the swelling remains weak ($R<50\,R_\odot$) for all the accretion histories considered.
The physical reason of this behaviour is that spherical accretion during the early accretion phase
leads to high stellar luminosities at the end of this phase, allowing efficient losses of entropy.
Then, disc accretion in the later phases lowers the amount of entropy that is accreted by the star.
Our results indicate that the combination of these two effects is able to reduce drastically the magnitude of the swelling,
with significant consequences on the evolution of the ionizing feedback and the size of the surrounding \hii\ region.
Interestingly, we notice that this scenario of early spherical accretion and later disc accretion
is the more realistic, since the formation of a disc in the accretion flow is not instantaneous.

\subsection{Comparison with other works}
\label{sec-dis-lit}

The pre-MS evolution of the models described above are in good agreement with those of \cite{haemmerle2016a},
obtained with the same stellar code in the idealized case of a constant accretion rate ($\dot M=10^{-4}-10^{-3}\,M_\odot\rm\,yr^{-1}$),
with the same assumption of cold disc accretion.
With the more realistic time-dependent accretion rates considered here, that remain essentially in the same range,
we obtain stellar structures and evolutionary tracks that are qualitatively similar to the case of a constant rate\footnote{\ 
   We notice that the CV and RD initial models in \cite{haemmerle2016a} are not exactly the same as those used here.}.
The key quantity for a quantitative comparison is the maximum radius $R_{\rm max}$ reached during the swelling.
In the CV case, \cite{haemmerle2016a} obtained $R_{\rm max}\simeq20\,R_\odot$ for $\dot M=10^{-4}\,M_\odot\rm\,yr^{-1}$,
and $R_{\rm max}\simeq40\,R_\odot$ for $\dot M=10^{-3}\,M_\odot\rm\,yr^{-1}$.
The value we obtain with our time-dependent accretion rates are in the same range.
In the RD case, \cite{haemmerle2016a} obtained $R_{\rm max}\simeq204\,R_\odot$ for $\dot M=10^{-3}\,M_\odot\rm\,yr^{-1}$.
With the lower rates used here, we remain under this value.
Moreover, in the models of \cite{haemmerle2016a}, the value of $R_{\rm max}$ increases with $\dot M$, as in our case.

\cite{hosokawa2009} and \cite{hosokawa2010} computed pre-MS models with accretion at constant rates,
based on both assumptions of spherical accretion and cold disc accretion.
In their fiducial cases, they considered $\dot M=10^{-3}\,M_\odot\rm\,yr^{-1}$, which is slightly higher than the values we use here,
and they used initial models of $\simeq0.1\,M_\odot$ that were fully radiative, having thus an entropy profile increasing outwards.
In their fiducial model for cold disc accretion (\citealt{hosokawa2010}), once the star reaches $2\,M_\odot$ by accretion, its radius is $3.5\,R_\odot$.
By comparing this value with the radius of our two initial models ($M=2\,M_\odot$, Sect.~\ref{sec-st-ini}),
we see that the fiducial model of \cite{hosokawa2010} corresponds to an intermediate case between our CV and RD cases,
although it is much closer to the RD case than to the CV case.
When the swelling occurs, the stellar radius reaches $R_{\rm max}\simeq100\,R_\odot$ in their model,
which is also an intermediate value between our CV and RD cases (resp. $44\,R_\odot$ and $198\,R_\odot$ for the accretion history 1).
In the case of spherical accretion, \cite{hosokawa2009} reached similar values for the maximum radius during the swelling.
It shows that the magnitude of the swelling does not change critically when we use purely spherical accretion instead of purely disc accretion.
However, as explained in Sect.~\ref{sec-dis-geo}, our results indicate that in the case of an accretion geometry
that is spherical in the early stages (low-mass range) and disc-like in the later stages (intermediate- and high-mass range),
the magnitude of the swelling is significantly reduced.
We notice that \cite{hosokawa2010} computed such models (labeled "MD3-SDm1").
In this case, they obtained only $R_{\rm max}\simeq50\,R_\odot$ (see their Fig.~10), which is consistent with our results.

\cite{kuiper2013} computed pre-MS evolution with more realistic accretion rates,
by coupling their stellar code with a hydrodynamic code for the accretion flow, and using the assumption of cold disc accretion.
The time-dependent accretion rates they obtained is again slightly higher than in our case,
with values of $\sim10^{-3}\,M_\odot\rm\,yr^{-1}$.
In all their models, when the star reaches a mass of $2\,M_\odot$, its radius is $\simeq1\,R_\odot$,
which is smaller than our two initial models, and than the models of \cite{hosokawa2010} at the same mass.
Thus, it is not surprising that the maximum radius reached during the swelling is much larger in their models than in ours:
in their run "RHD+SE A", the radius reaches $885\,R_\odot$, which is a factor $\sim4$ higher than in our case (accretion history 1).

All these examples show that our results are in good agreement with those of other authors,
provided that we compare models with the same physical conditions (initial model, accretion rate and accretion geometry).
Moreover, it confirms that the choice of the initial model (i.e. the early accretion geometry)
has critical consequences on the magnitude of the swelling, and then on the ionizing effect of the stellar radiation:
the more compact is the initial model, the stronger are the swelling and the reduction of the number of ionizing photons during the swelling.
Indeed, all the models with disc accretion presented in \cite{hosokawa2010} and \cite{kuiper2013} consider initial models
that are relatively compact (i.e. they model a star formed by disc accretion since the very beginning of the accretion phase),
leading thus to a strong swelling, like in our RD models,
in contrast to our CV models that are initially more extended and that have a much weaker swelling.
In particular, all these results suggest that a change in the accretion geometry in the early accretion phase only
(i.e. before the swelling, while the star is still in the low-mass range), has a much stronger impact than a global change:
the compact initial models of \cite{kuiper2013} and our extended CV initial models lead to values of $R_{\rm max}$
that differ by more than one order of magnitude, despite the fact that they are based on the same assumption of cold disc accretion
during the swelling phase;
while in the models of \cite{hosokawa2009} and \cite{hosokawa2010} that use various (but fixed) accretion geometry
during the whole accretion phase, the value of $R_{\rm max}$ remains in the same range.

\quad

\cite{klassen2012b} estimated the evolution of the ionizing flux and of the size of an \hii\ region produced by this flux during the swelling,
using the pre-computed stellar models of \cite{hosokawa2009}, for spherical accretion at constant rates ($10^{-5}-10^{-3}\,M_\odot\rm\,yr^{-1}$).
As mentioned above, these models are computed from compact radiative initial models, similar to our RD models,
and thus they experience a relatively strong swelling ($R_{\rm max}\simeq100\,R_\odot$ in the case $\dot M=10^{-3}\,M_\odot\rm\,yr^{-1}$).
In this case, the effective temperature drops from $10^4\,K$ to $6\times10^3\,K$ during the swelling,
and it results in a decrease in $S^*$ by 5 orders of magnitude in a timescale of 3500 years.
Such a decrease is consistent with our RD model in the case of accretion history 4.
But the assumption of spherical accretion in the models of \cite{hosokawa2009}
reflects in higher luminosities and effective temperatures during the swelling than in our models.
As a consequence, the ionizing flux is stronger in the models of \cite{hosokawa2009} than in ours, already before the swelling.
However, we notice that the typical magnitude of the decrease in $S^*$, and more importantly the timescale of this decrease,
are similar in our case than in the case of \cite{klassen2012b}.

\subsection{Comparison with observations}

\citet{mottram2011} have obtained luminosity functions of massive young stellar objects (MYSOs)
and compact \hii\ regions for the Red \emph{MSX} Source (RMS) survey. They find a lack of high-luminosity
MYSOs and speculate that these MYSOs may be too cool to ionize their surroundings because they are undergoing a swelling phase
due to high accretion rates. However, the drop in effective temperature during the swelling is in general a relatively short-lived phase compared
to the persistently lower effective temperatures in our CV models, as it is visible on Fig.~\ref{fig-ion1} to \ref{fig-ion7}.
This suggests that convective initial conditions during high-mass pre-MS evolution may be a more compelling solution to the discrepancy,
but we leave a more thorough analysis for future work.

\section{Conclusions}
\label{sec-ccl}

We have computed pre-MS tracks for massive and intermediate-mass stars with variable accretion rates,
using the assumption of cold disc accretion.
We have started our models from two different initial conditions (at $M=2\,M_\odot$):
the fully convective model CV and the fully radiative model RD,
that can be seen respectively as the product of spherical and disc accretion in the early phase
(i.e. in the low-mass range $M<2\,M_\odot$).

We find significant differences between the two cases for the evolution of the stellar radius and the ionizing luminosity.
In general, the CV case shows a much weaker swelling of the protostar for high accretion rates than the RD case.
It indicates that in the case of an accretion geometry that is spherical in the early accretion phase (before the swelling)
and disc-like in the later accretion phase (during the swelling),
the magnitude of the swelling is drastically reduced compared to the case of a purely disc geometry,
with significant impact on the ionizing feedback of the star and the size of the surrounding \hii\ region.
The stellar structure in case RD is very sensitive to the accretion rate and reacts quickly to accretion bursts,
which can lead to considerable changes in stellar radius, effective temperature and ionizing luminosity
on timescales as short as 100 -- 1000 yr.
In contrast, the evolution in case CV is much less influenced by the instantaneous accretion rate,
and always produces a monotonically increasing ionizing photon flux that can be many orders of magnitude smaller
than the corresponding flux in case RD.
For massive stars, this difference in the ionizing flux results in a delay of the \hii\ region expansion by up to 10,000 yr.
While in the CV case the radius of the \hii\ region is already 100\,--\,1000\,AU when the stellar luminosity reaches $10^4\,L_\odot$,
in the RD case it remains as low as 1\,AU at such high luminosities.
This last value is similar to the stellar radius itself in this case, due to the strong swelling,
so that the \hii\ region disappears nearly or completely, which never happens in the CV case.

We conclude that the choice of the initial protostellar model (i.e. of the early accretion geometry)
has a large impact on the strength of the radiative feedback in general, and the ionizing luminosity in particular,
during the pre-MS evolution of massive protostars, with significant consequences on the size of \hii\ regions around them.
This effect is much larger than the difference between purely spherical and purely disc accretion,
and introduces an important uncertainty that should be taken into account in models of high-mass star formation.
Interestingly, because of their lower effective temperatures, our CV models may hint at a solution to an observed discrepancy
between the luminosity distribution functions of MYSOs and compact \hii\ regions.

\appendix

\section{Initial size of the radiative core and magnitude of the swelling}
\label{sec-rc}

\begin{figure}\includegraphics[width=0.49\textwidth]{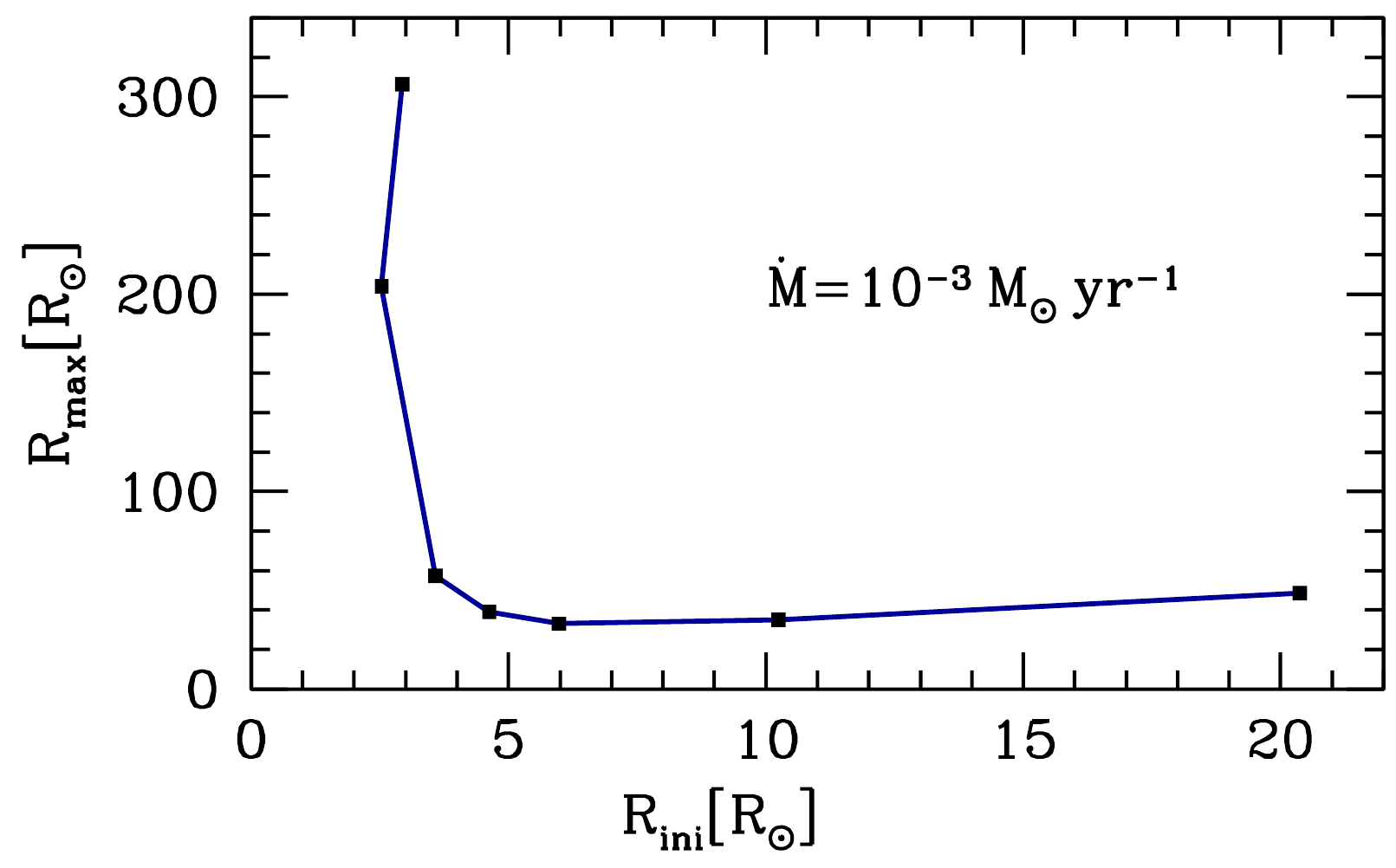}
\caption{Maximum radius reached during the swelling as a function of the initial radius,
for the seven models described in Sect.~\ref{sec-rc} (Table~\ref{tab:ini}).}
\label{fig-inir}\end{figure}

\begin{figure}\includegraphics[width=0.49\textwidth]{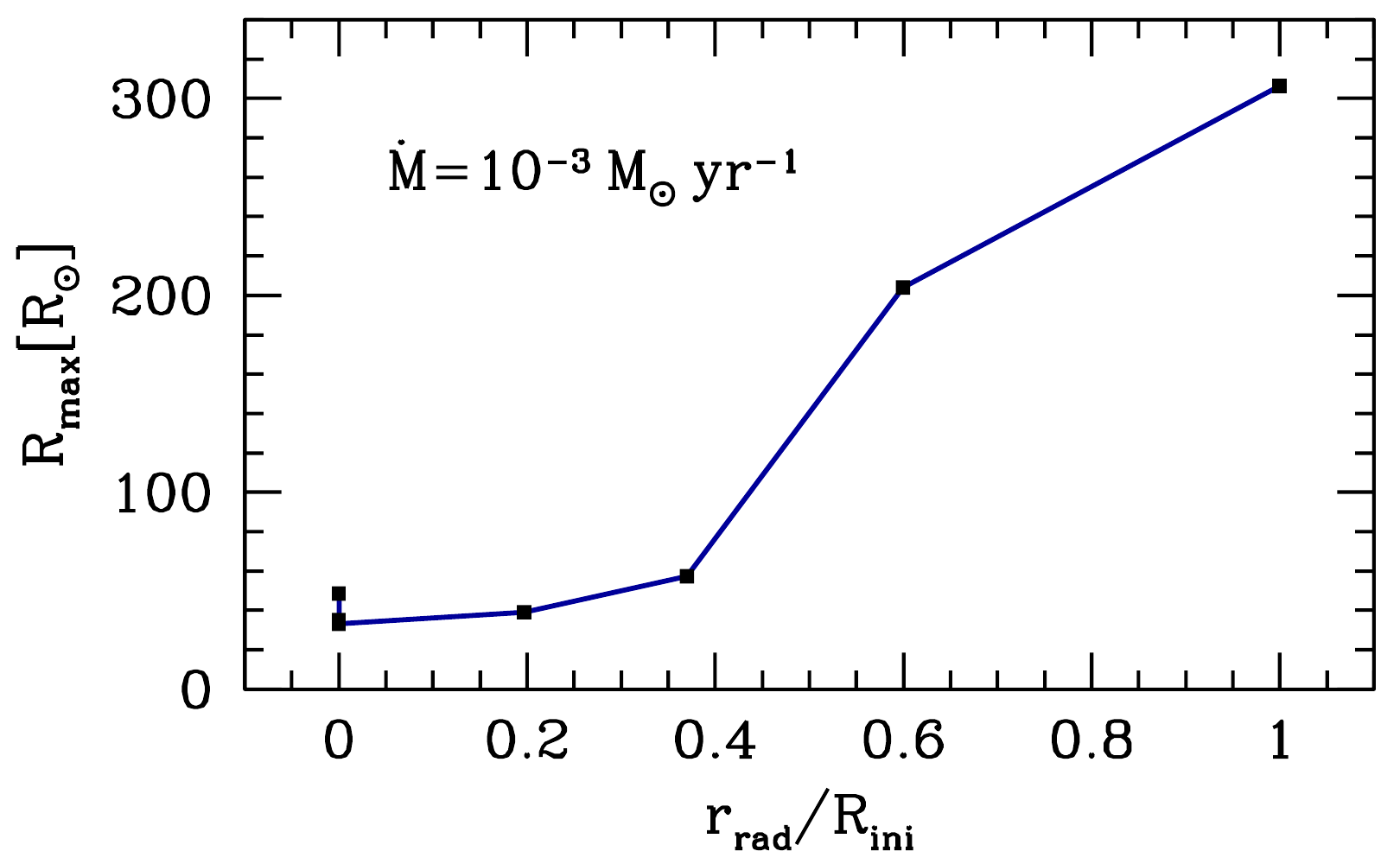}
\caption{Maximum radius reached during the swelling for the seven models described in Sect.~\ref{sec-rc} (Table~\ref{tab:ini})
as a function of $r_{\rm rad}/R_{\rm ini}$, where $r_{\rm rad}$ is the radius of the radiative core of the initial model
and $R_{\rm ini}$ is the total radius of the initial model.}
\label{fig-inirc}\end{figure}

The models described above were computed using two initial models, CV and RD,
being respectively fully convective and fully radiative (see Sect.~\ref{sec-st-ini}).
In the present appendix, we consider several intermediate initial models and we look at the evolutionary tracks
obtained from these initial conditions in the simplified case of a constant accretion rate of $10^{-3}\,M_\odot\rm\,yr^{-1}$.
We use seven initial models that differ from each other in their total radii as well as their internal structure:
three of them are fully convective, three are made of a radiative core and a convective envelope, and the last one is fully radiative.
Table \ref{tab:ini} gives the values of the total radii $R_{\rm ini}$ and the radii $r_{\rm rad}$ of their radiative core as a percentage of $R_{\rm ini}$.
It shows also the maximum radius $R_{\rm max}$ reached during the swelling.
The value of $R_{\rm max}$ as a function of $R_{\rm ini}$ is shown on Fig.~\ref{fig-inir},
and as a function of $r_{\rm rad}/R_{\rm ini}$ on Fig.~\ref{fig-inirc}.
Notice that the two extreme models correspond to our CV and RD models.

\begin{table}
\caption{Properties of the seven models described in App.~\ref{sec-rc}.}
\label{tab:ini}
\centering
\begin{tabular}{c c c}
\hline
\hline
$R_{\rm ini}$ ($R_\odot$) & $r_{\rm rad}/R_{\rm ini}$ & $R_{\rm max}$ ($R_\odot$)\\
\hline
20.4 & 0 & 48.6 \\
10.3 & 0 & 35.0 \\
6.0 & 0 & 33.1 \\
4.6 & 20\% & 39.2 \\
3.6 & 37\% & 57.5 \\
2.5 & 60\% & 204 \\
2.9 & 100\% & 306 \\
\hline
\end{tabular}
\end{table}

As we see, the value of $R_{\rm max}$ for the most compact models, which are not fully convective,
is much larger than for the extended fully convective models.
However, we notice that the relation between $R_{\rm max}$ and $R_{\rm ini}$ is not monotonic.
Indeed, by comparing the three models that are initially fully convective, we see that a strong decrease in $R_{\rm ini}$,
from 20.4 to $6\,R_\odot$, reduces slightly the value of $R_{\rm max}$.
But for the models that have initially a radiative core, even a modest decrease in $R_{\rm ini}$,
from 4.6 to $2.5\,R_\odot$, leads to a spectacular increase of $R_{\rm max}$.
By comparing the two models that are the more compact, we even see that in this case the one with a slightly higher $R_{\rm ini}$
(namely the RD model) leads to a much larger $R_{\rm max}$.
If we look instead at the relation between $R_{\rm max}$ and $r_{\rm rad}/R_{\rm ini}$,
we see that the correlation is much more straightforward: the value of $R_{\rm max}$ changes only slightly between the models
that have $r_{\rm rad}=0$, despite the significant difference between their total radii,
and it starts to increase significantly only when $r_{\rm rad}$ increases too.
This result indicates that the significant quantity that determines the magnitude of the swelling is not the stellar radius,
but the size of the radiative core.

\section{Impact of deuterium shell-burning on the swelling}
\label{sec-deut}

\begin{figure}\includegraphics[width=0.49\textwidth]{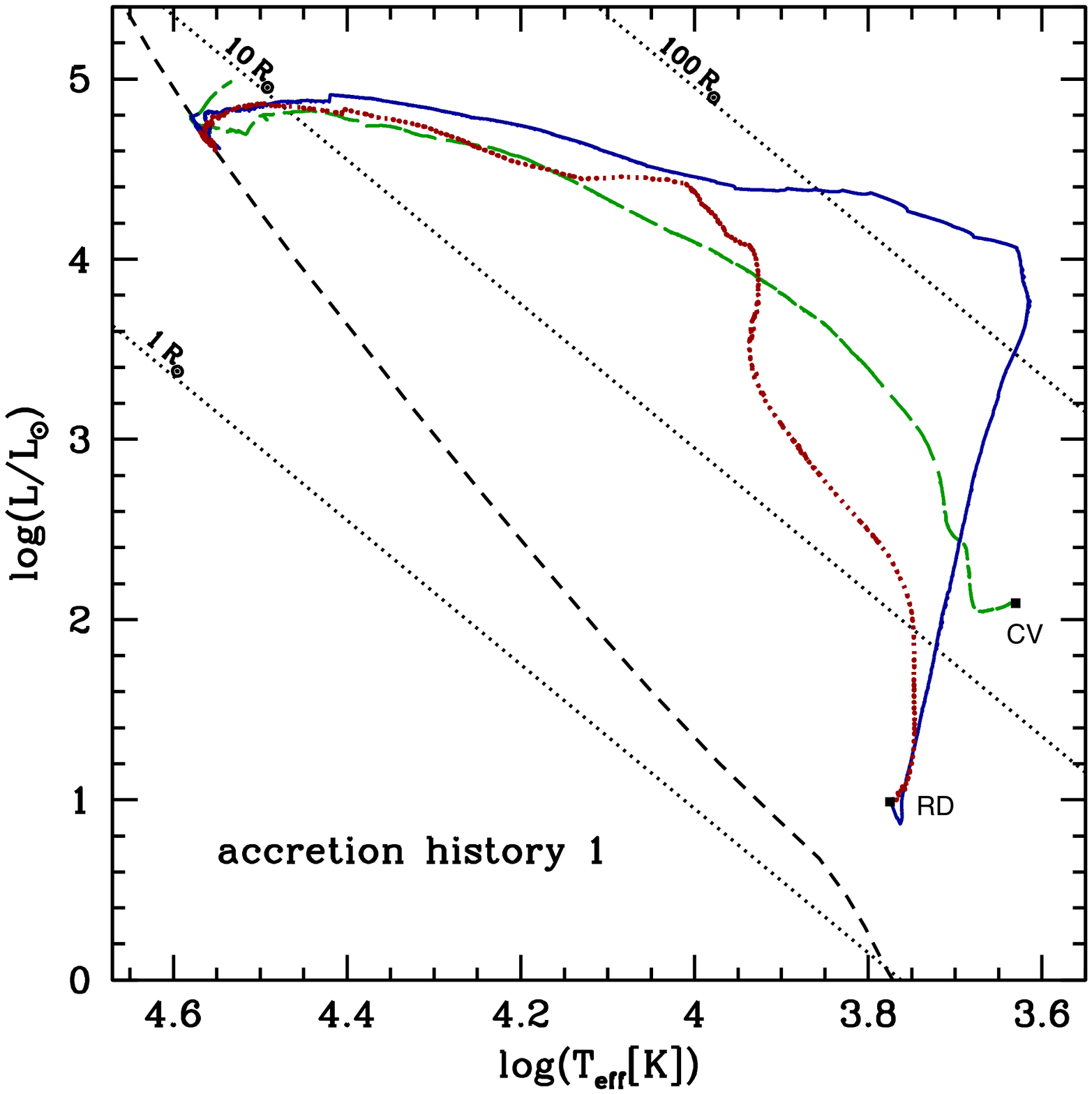}
\caption{Evolutionary tracks of the three models computed with accretion history 1
(described in Sect.~\ref{sec-st-res-1} and \ref{sec-deut}):
the case with the initial model CV, that includes deuterium, is indicated by the green dashed line;
the case with the initial model RD including deuterium in the accreted material is indicated by the solid blue line;
the case with the initial model RD without any deuterium is indicated by the red dotted line.}
\label{fig-hrD0_1}\end{figure}

\begin{figure}\includegraphics[width=0.49\textwidth]{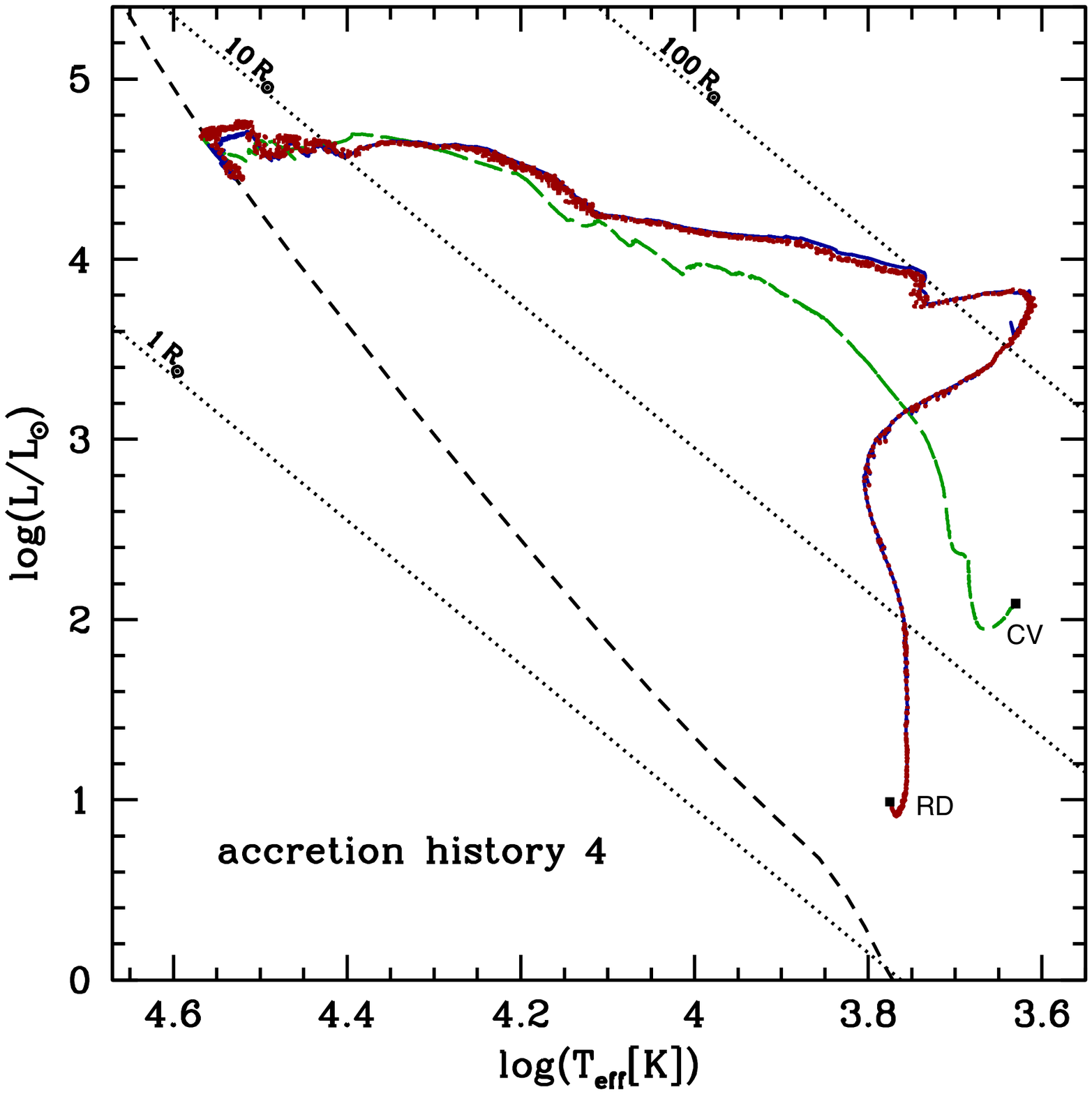}
\caption{Same as Fig.~\ref{fig-hrD0_1} but for accretion history 4.}
\label{fig-hrD0_4}\end{figure}

\begin{figure}\includegraphics[width=0.49\textwidth]{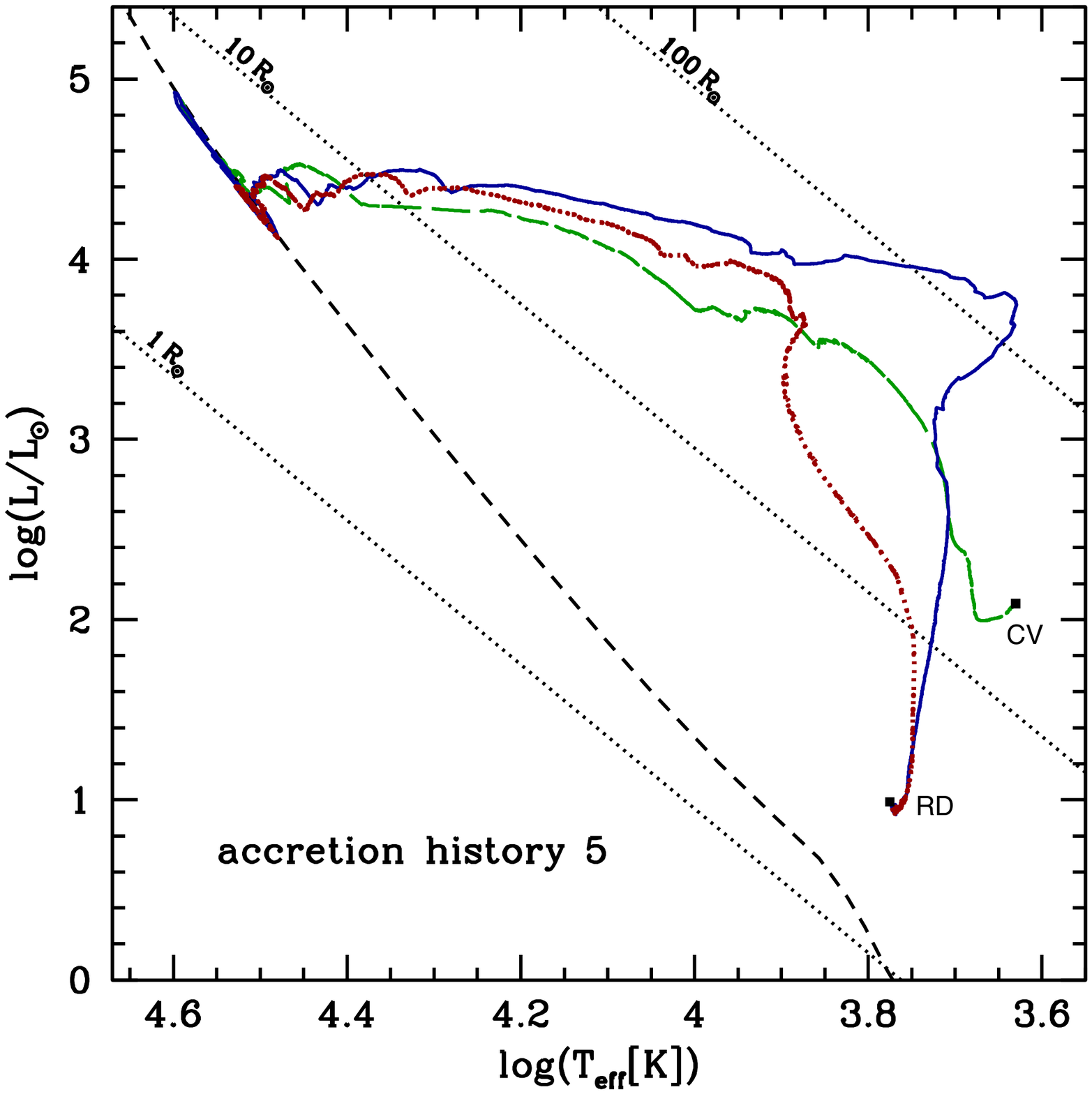}
\caption{Same as Fig.~\ref{fig-hrD0_1} but for accretion history 5.}
\label{fig-hrD0_5}\end{figure}

As mentioned in Sect.~\ref{sec-st-res-1}, the entropy release that produces the swelling comes
from both gravitational contraction and deuterium shell-burning.
In order to evaluate the relative contribution of these two entropy sources,
we reconsider here the cases that lead to the strongest swelling (i.e. accretion histories 1, 4 and 5, with the RD initial model),
using this time $X_2=0$, i.e. without including any deuterium in the accreted material.
The resulting evolutionary tracks are plotted on Figs.~\ref{fig-hrD0_1}, \ref{fig-hrD0_4} and \ref{fig-hrD0_5} (red dotted tracks)
together with the evolutionary tracks of the equivalent models that include deuterium, already described in Sect.~\ref{sec-st-res}.

The effect of deuterium shell-burning is visible by comparing the red and the blue tracks on the three plots.
As we see, the relative importance of D-burning depends sensitively on the accretion history.
For accretion histories 1 and 5, removing the deuterium in the accreted material reduces significantly the magnitude of the swelling.
Without deuterium, the maximum radius reached during the swelling
is similar to the CV case with deuterium ($R_{\rm max}\simeq50\,R_\odot$).
In contrast, for accretion history 4, the impact of D-burning on the magnitude of the swelling is negligible:
the red and blue curves nearly overlap.
We notice that for accretion histories 1 and 5, the accretion rate reaches values $>5\times10^{-4}\,M_\odot\rm\,yr^{-1}$
in the very beginning of the swelling, before decreasing slowly,
while for accretion history 4 such values are reached later, once the swelling is already significant.
It suggests that the effect of D shell-burning dominates when a star having an RD structure
accretes at a high rate ($\sim10^{-3}\,M_\odot\rm\,yr^{-1}$).
But when the accretion rate on such a star is low ($\sim10^{-4}\,M_\odot\rm\,yr^{-1}$),
D-burning plays a negligible role, even if the rate grows once the swelling is already significant.

Since the initial model RD is built without including any deuterium (because of its high internal temperature, see Sect.~\ref{sec-st-ini}),
the effect of D-burning in the models computed from it is only due to the deuterium that is accreted during the evolution.
Thus one can wonder if this effect is real or artificial, due to the addition of D in a model that did not contain initially such a chemical species,
and to the sudden triggering of its combustion.
Indeed, a comprehensive answer to this issue would require to start the computations at lower masses,
when the internal temperature is still too low for the burning of deuterium, and to follow consistently its destruction.
Unfortunately, such computations are currently problematic with the Geneva code,
due to difficulties in the convergence process at low masses when we include accretion.
Nevertheless, we were able to compute one model starting from a mass of $0.3\,M_\odot$,
accreting at a high constant rate of $10^{-3}\,M_\odot\rm\,yr^{-1}$ (let us call this model \texttt{ini03}).
When this model reaches $M=2\,M_\odot$, its internal structure and surface properties
correspond to an intermediate case between our CV and RD models.
If we build such an intermediate model (let us call it RC\footnote{\ 
   This RC model is the same as the one with $R_{\rm ini}=3.6\,R_\odot$ described in App.~\ref{sec-rc} (Table~\ref{tab:ini}).})
in the same way as we built the RD model,
we obtain a central temperature of $4\times10^6\,K$, which is still much higher than the temperature required for significant D-burning,
and we have to take $X_2=0$ in the whole star, like in the RD case.
Then we can compute the evolution starting from RC, at the same constant accretion rate of $10^{-3}\,M_\odot\rm\,yr^{-1}$,
in both cases with and without deuterium in the accreted material, and compare it with the evolution of \texttt{ini03}.

\begin{figure}\includegraphics[width=0.49\textwidth]{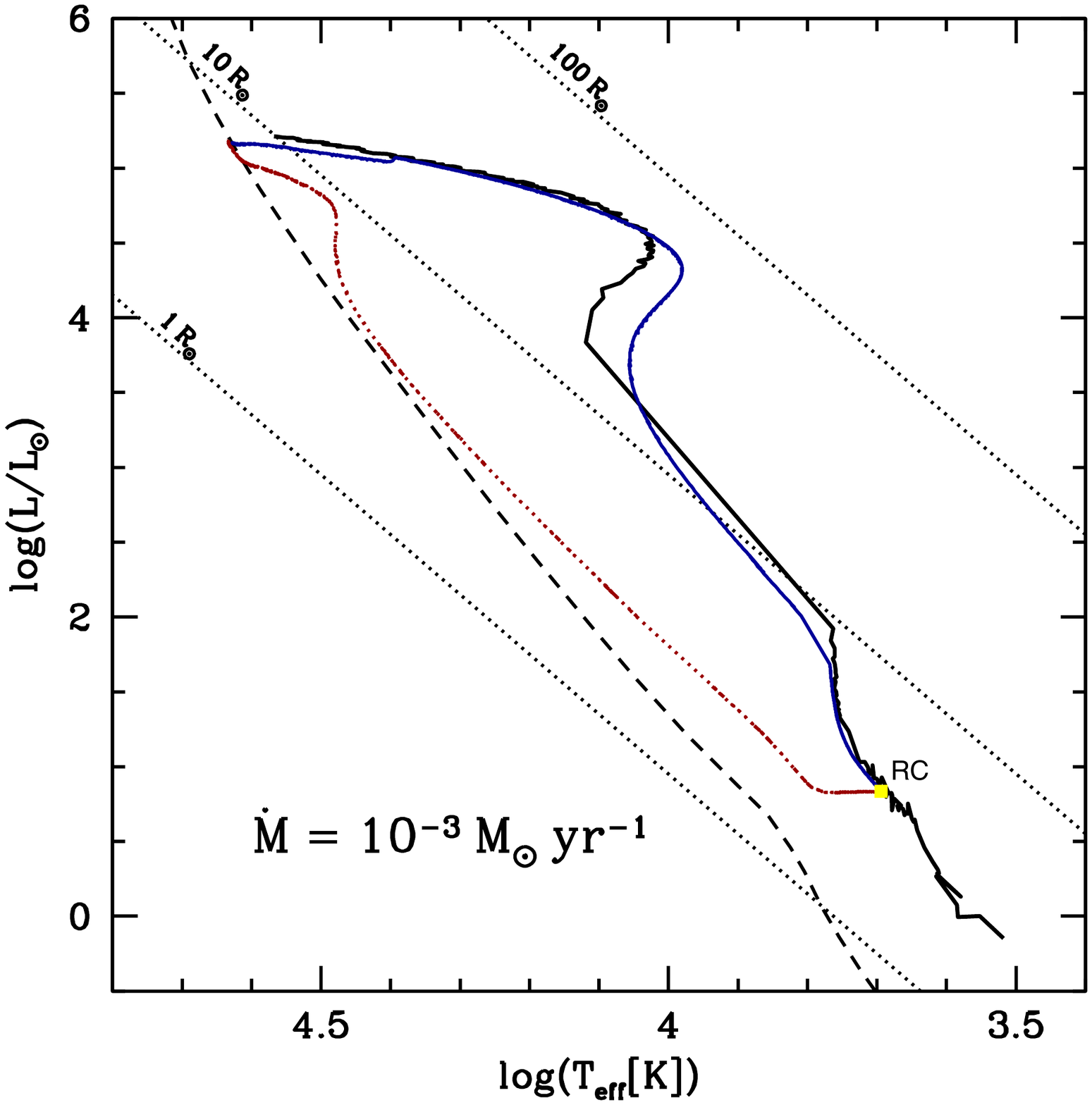}
\caption{Evolutionary tracks of the models accreting at the constant rate $10^{-3}\,M_\odot\rm\,yr^{-1}$
starting from the initial models \texttt{ini03} (solid black line) and RC described in App.~\ref{sec-deut}.
The solid blue line is the RC model with deuterium in the accreted material,
and the red dotted line is the RC model without deuterium.
The black dashed curve is the ZAMS \citep{ekstroem2012},
and the black dotted straight lines are iso-radius of 1, 10 and $100\,R_\odot$.}
\label{fig-ini03}\end{figure}

The evolutionary tracks of these models are shown on Fig.~\ref{fig-ini03}.
As we see, the blue track reproduces well the black one, i.e. the RC initial model leads to a pre-MS evolution
which is similar to a model computed from a lower initial mass, in which we follow consistently the whole D-burning phase.
In contrast, the red track diverges quickly from the other two, showing that in this case the deuterium shell-burning
plays the dominant role in the swelling.

This example confirms on one hand the importance of D-burning, but it shows on the other hand
that the effect of D-burning is well reproduced with the initial model RC,
despite the fact that this initial model does not contain any deuterium.
We conclude that the impact of D-burning on the magnitude of the swelling is dominated by the burning of the accreted deuterium.
Even if some deuterium survives to the early accretion phase ($M<2\,M_\odot$), it plays a negligible role in the subsequent evolution.

\section{Impact of initial conditions on the swelling in the case of a constant accretion timescale}
\label{sec-taccr}

In Sect.~\ref{sec-dis-sw}, we interpret the dependence of the magnitude of the swelling on the initial conditions
by comparing the Kelvin-Helmholtz time of our initial models.
However, we notice that the mass at which the swelling occurs depends also on the initial conditions,
so that the accretion timescale at the beginning of the swelling differs between the CV and the RD cases.
In order to evaluate how this feature influences our results, we compute models starting from the same CV and RD initial models,
but using an accretion rate $\dot M\propto M$, for which $t_{\rm accr}$ is constant.
We take $t_{\rm accr}=10^4\rm\,yr$, i.e. $\dot M=M/10^4\rm\,yr$.
This rate has the same order of magnitude as the time-variable rates used in Sect.~\ref{sec-st-res},
running from $10^{-4}\,M_\odot\rm\,yr^{-1}$ for $M=1\,M_\odot$ to $10^{-3}\,M_\odot\rm\,yr^{-1}$ for $M=10\,M_\odot$.

\begin{figure}\includegraphics[width=0.49\textwidth]{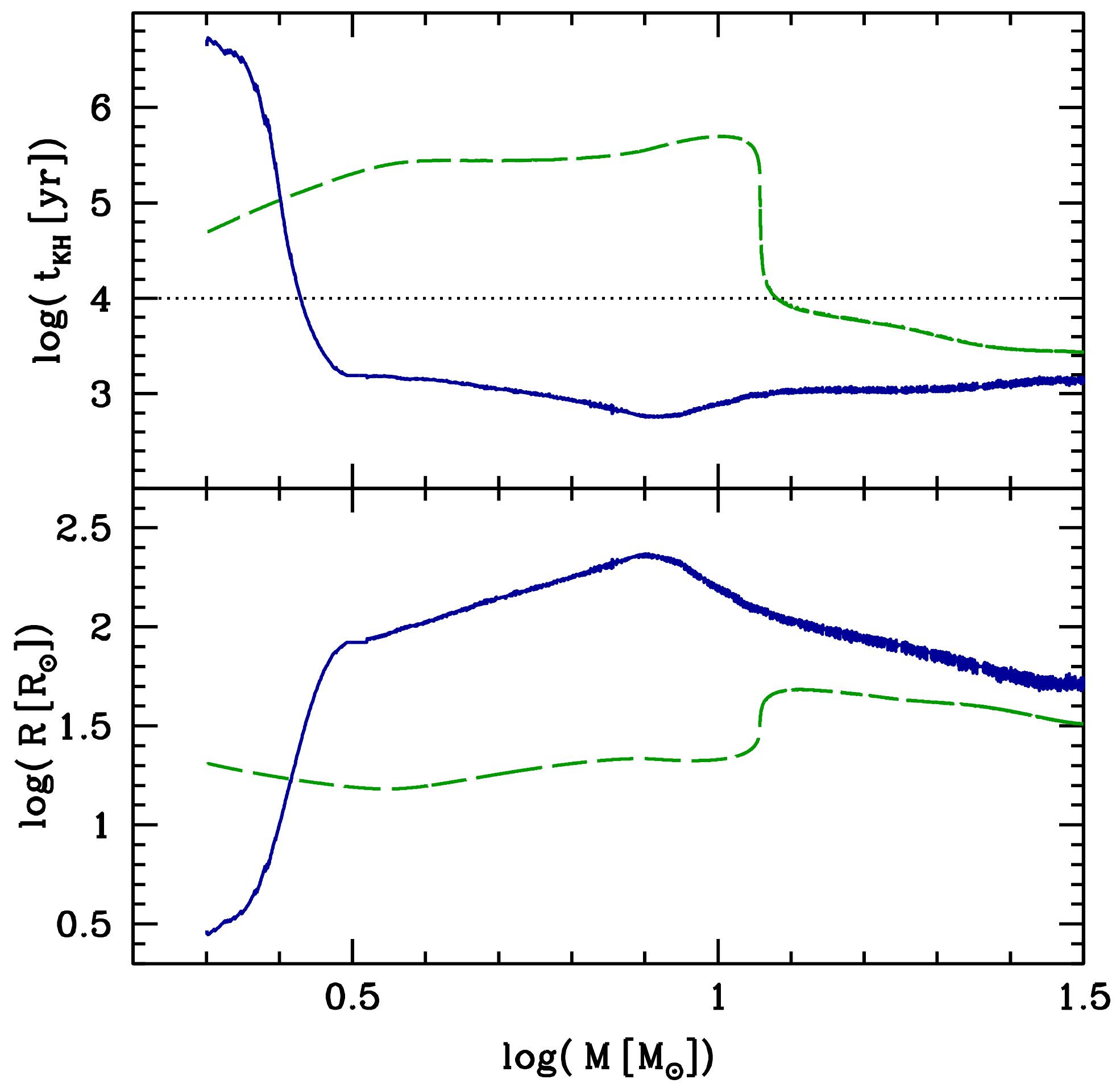}
\caption{Evolution of the Kelvin-Helmholtz time (upper panel) and the stellar radius (lower panel)
for the models described in App.~\ref{sec-taccr}.
On each panel, the RD model is indicated by a solid blue line, and the CV model by a green dashed line.
The horizontal black dotted line on the upper panel indicates the accretion timescale $t_{\rm accr}=10^4\rm\,yr$.}
\label{fig-taccr}\end{figure}

The evolution of the timescales and the radii of these models is shown on Fig.~\ref{fig-taccr}.
As for the models described in Sect.~\ref{sec-st-res} and \ref{sec-dis-sw},
the evolution starts with $t_{\rm KH}({\rm RD})>>t_{\rm KH}({\rm CV})>>t_{\rm accr}$.
In the RD case, the swelling starts since the beginning of the evolution ($M=2\,M_\odot$), $t_{\rm KH}$ decreases,
becomes rapidly shorter than $t_{\rm accr}$, and the radius exceeds $100\,R_\odot$, reaching $R_{\rm max}=234\,R_\odot$.
In contrast, the CV model starts its evolution by the slow central D-burning phase, and the swelling occurs only at a mass of $\simeq10\,M_\odot$.
But despite the fact that the accretion timescale remains constant, the swelling is still much weaker than in the RD case,
leading to a maximum radius of $48\,R_\odot$ only.

We conclude that the increase in the accretion timescale with the mass in the models described in Sect.~\ref{sec-st-res}
and \ref{sec-dis-sw} is not responsible for the difference in the magnitude of the swelling between the CV and the RD cases.

\section*{Acknowledgements}

We are grateful to the referee for their useful remarks and questions, which helped us to improve the paper significantly.
T.~P. acknowledges support from the DFG Priority Program 1573 {\em Physics of the Interstellar Medium}.
Part of this work was supported by the Swiss National Science Foundation.




\bibliographystyle{mn2e}
\bibliography{bibliotheque} 




%
%


\bsp	
\label{lastpage}
\end{document}